\PassOptionsToPackage{usenames,dvipsnames}{xcolor}
\documentclass[a4paper,11pt,1p]{elsarticle}
\usepackage[titletoc,toc,title]{appendix}

\usepackage[T1]{fontenc}
\usepackage[utf8]{inputenc}
\usepackage[british]{babel}
\addto\captionsbritish{}
\usepackage{slashbox} 
\usepackage{geometry}
\usepackage{graphicx} 
\usepackage{amssymb} 
\usepackage{amsmath,pdftexcmds}
\usepackage{fancyvrb}
\usepackage{multicol}
\usepackage{multirow}
\usepackage{mathtools}
\usepackage{bbm}
\usepackage{amsbsy}
\usepackage{tabularx}
\usepackage{hyperref}
\usepackage{adjustbox}  
\usepackage{url}
\usepackage{booktabs}
\usepackage{mathrsfs}
\usepackage{pgffor}
\usepackage{textcomp}
\usepackage[version=3]{mhchem}
\usepackage{framed}
\usepackage{pgfplots}
\usepackage{filecontents}
\usepackage{comment}
\usepackage{cuted} 
\usepackage{youngtab}
\usepackage{tikz}
\usepackage{caption}
\usepackage[all]{xy}
\usepackage{amsfonts}
\usepackage{anysize}
\marginsize{2.4cm}{2.4cm}{1.2cm}{2.6cm}

\DeclareFontFamily{U}{cbgreek}{}
\DeclareFontShape{U}{cbgreek}{m}{n}{
        <-6>    grmn0500
        <6-7>   grmn0600
        <7-8>   grmn0700
        <8-9>   grmn0800
        <9-10>  grmn0900
        <10-12> grmn1000
        <12-17> grmn1200
        <17->   grmn1728
      }{}
\DeclareFontShape{U}{cbgreek}{bx}{n}{
        <-6>    grxn0500
        <6-7>   grxn0600
        <7-8>   grxn0700
        <8-9>   grxn0800
        <9-10>  grxn0900
        <10-12> grxn1000
        <12-17> grxn1200
        <17->   grxn1728
      }{}

\DeclareRobustCommand{\digamma}{%
  \text{\usefont{U}{cbgreek}{\normalorbold}{n}\symbol{147}}%
}

\newcommand*\bfell{\ensuremath{\boldsymbol\ell}}

\makeatletter
\newcommand{\normalorbold}{%
  \ifnum\pdf@strcmp{\math@version}{bold}=\z@ bx\else m\fi
}
\makeatother

\makeatletter
\def\ps@pprintTitle{
   \let\@oddhead\@empty
   \let\@evenhead\@empty
   \def\@oddfoot{\reset@font\hfil\thepage\hfil}
   \let\@evenfoot\@oddfoot
}
\makeatother

\newsavebox{\leftbox}
\newsavebox{\rightbox}

\definecolor{DarkMidnightBlue}{rgb}{0.0, 0.04, 0.14}
\definecolor{DarkOrchid}{rgb}{0.6, 0.2, 0.8}
\definecolor{BrandeisBlue}{rgb}{0.0, 0.44, 1.0}

\renewcommand{\baselinestretch}{1.25} 

\hypersetup{hidelinks,
backref=true,
pagebackref=true,
hyperindex=true,
breaklinks=true,
colorlinks=true,
linkcolor=blue, 
citecolor=blue, 
urlcolor=DarkMidnightBlue,
bookmarks=true,
bookmarksopen=false,
pdftitle={Title},
pdfauthor={Author}}

\VerbatimFootnotes

\title{\textsf{\textbf{Magnetic dipole moments as a strong signature for $\alpha$-clustering in even-even self-conjugate nuclei}}}

\author[hiskp,bctp,cea,esnt]{Gianluca~Stellin}

\author[hiskp]{Karl-Heinz~Speidel}

\author[hiskp,bctp,fzj,tsu]{Ulf-G.~Mei\ss{}ner}

\address[hiskp]{Helmholtz Institut f\"ur Strahlen- und Kernphysik, Universit\"at Bonn, Nu\ss{}allee 14-16,
53115 Bonn, Germany}
\address[bctp]{Bethe Center for Theoretical Physics, Universit\"at Bonn, Nu\ss{}allee 12, 53115 Bonn, Germany}
\address[cea]{DRF/IRFU/DPhN/LENA, CEA Paris-Saclay, Bât. 703, 91190 Saint-Aubin, France}
\address[esnt]{Espace de Structure Nucleaire Théorique, CEA Paris-Saclay, Bât. 703, 91190 Saint-Aubin, France}
\address[fzj]{Institute for Advanced Simulation, Institut f\"ur Kernphysik and J\"ulich Center for Hadron Physics, \\
Forschungszentrum J\"ulich, 52425 J\"ulich, Germany}
\address[tsu]{Ivane Javakhishvili Tbilisi State University,  0186 Tbilisi, Georgia}

\date{\today}

\begin{document}


\begin{abstract}
\begin{small}
We investigate the magnetic dipole moments in even-even self-conjugate nuclei from
\ce{^{12}C} to \ce{^{44}Ti}. For the latter, the measured gyromagnetic factors of excited states turn out to assume the same value of $g \approx + 0.5$ 
within  statistical errors. This peculiar feature can be interpreted on the basis of collective excitations of  $\alpha$-clusters. Analogously, the
behaviour of the same observable is studied for all isotopes obtained by adding one or two neutrons to the considered self-conjugate nuclei. 
It is found that for the $N = Z + 1$ isotopes the $\alpha$-cluster structure hardly contributes to the observed negative g- factor value, corroborating 
molecular $\alpha$-cluster models. The addition of a further neutron, however, restores the original $\alpha$-cluster g-factors, except 
for the semi-magic isotopes, in which the deviations from $g \approx + 0.5$ can be associated with the relevant shell closures. 
Secondly, we analyze the same observable in the framework of a macroscopic $\alpha$-cluster model on a finite lattice of side length $L$. We focus on the discretization effects induced in the magnetic dipole moments of the $2_1^+$ and the $3_1^-$ states of \ce{^{12}C} at different values of the lattice spacing $a$. 
\end{small}
\end{abstract}

\begin{keyword}
Electromagnetic Procs. and Props. \sep Nucl. Struct. Mod. and Meth. \sep Few-Body Syst.
\PACS 12.38.Gc \sep 21.60.-n \sep 25.45.-v
\end{keyword}

\maketitle


\tableofcontents

\clearpage


\section{\textsf{Preamble}}\label{S-1.0}

\hspace{\parindent} The tendency of nucleons to congregate into clusters is known since the early days of nuclear physics \cite{Whe37}. The intrinsic stability 
of the \ce{^4He} nucleus alongside with the energy gap of $20.21$~MeV with respect to the lowest single-particle excited state, makes the $\alpha$-particle a candidate
for a conglomerate of nucleons capable of surviving relatively unperturbed within the nuclear mean field,
for a recent review see \cite{FHK18}. It is in fact  well established that nuclear
spectra are a result of the interplay between pairing and cluster correlations and the mean field generated by the individual nucleons \cite{KaE19}.
In absence of correlations between the constituents, nucleons would move as independent particles and a well-pronounced shell
structure (the Fermi surface) would appear \cite{KaE19}. As soon as the residual attractive interactions are turned on, nucleons orbiting in the same subshell but 
with opposite projection of the angular momentum along the symmetry axis of the nucleus (i.e. connected by the antiunitary time-reversal operation, 
see refs.~\cite{SDB11,SBD13}) form Cooper pairs \cite{Coo56,Sol58} or give rise to larger subsystems. Among the latter, the $\alpha$-particle 
dominates in nuclei with even and equal numbers of protons and neutrons, whose higher binding energy per nucleon reaches a local maximum
 in the Segrè chart. Cluster correlations are found to reduce the neutron skin thickness with respect to a mean field value \cite{Typ14}. 

Additionally, in the pioneering work in ref.~\cite{HaT38} it was noticed that the binding energies of light $\alpha$-conjugate nuclei scale linearly with the
number of bonds among the $\alpha$-clusters, sitting at the vertices of regular polyhedra (cf. fig.~1 in ref.~\cite{BrB67} and refs.~\cite{LaM14,BiI20,HaR20}).  In particular, the adoption of the equilateral 
triangle ($\mathcal{D}_{3h}$) \cite{SFV16,FSV17,VCF20}, the tetrahedron ($\mathcal{T}_d$) and the triangular bipyramid ($\mathcal{D}_{3h}$) as basis-configuration for the 
inspection of the structure properties of the \ce{^{12}C} \cite{BiI02} and \ce{^{16}O} \cite{BiI14,BiI17} within the Algebraic Cluster Model (ACM) \cite{BiI20,BiI00} and of the \ce{^{20}Ne} 
\cite{BiI21-01} within a geometric macroscopic $\alpha$-cluster model, respectively, has revived the interest for the subject in recent times (cf. refs.~\cite{Jen16,CFL21}). 

Furthermore, $\alpha$-clustering is reinforced as the excitation energy grows. As an example, in \ce{^{12}C} at around $10$~MeV the nuclear 
density reduces to less than one third with respect to the initial value (cf. refs.~\cite{CFL21}) and the inter-$\alpha$ separation increases accordingly. 
An example is represented by the Hoyle state at $7.65$~MeV, credited by miscroscopic studies as a triangular \textit{bent-arm} arrangement of $\alpha$-particles \cite{EKL12,SLL22} and earlier as 
a linear chain configuration \cite{Mor56}. On the other hand, the triangular ground state keeps a significant overlap with the $p_{3/2}$ closed subshell configuration (cit. refs.~\cite{KaE19}), as the 
independent-particle in a mean field feature gains ground. In the opposite limit, the nucleus dissociates completely into the twelve constituents at 
around $10^2$~MeV \cite{KaE19}.

The idea according to which the $\alpha$-particle structure is not manifest in the ground state but emerges gradually with the increase of the internal energy 
of the system is at the origin of the diagrammatic representation proposed in refs.~\cite{ITH68,IHS80} for the \ce{^{12}C}, \ce{^{16}O}, \ce{^{20}Ne},
\ce{^{24}Mg} and \ce{^{28}Si} nuclei. The ensuing \textit{Ikeda diagram} predicts fully-clustered states, i.e. $M\alpha$-cluster configurations with $2M = Z = N$, 
only in the correspondence of the $M\alpha$-breakup thresholds.

Therefore, the adoption of pictures that neglect the inner degrees of freedom of the \ce{^4He} clusters for the analysis of excited state properties, such as intrinsic 
magnetic dipole moments, of this class of nuclei appears justified. Motivated by the long standing experimental interest in magnetic
moments for \ce{^{12}C} \cite{KHK79}, \ce{^{16}O} \cite{ABD84}, \ce{^{20}Ne} \cite{LSK03}, \ce{^{24}Mg} \cite{KSG15-02}, \ce{^{28}Si} \cite{SGH75}, \ce{^{32}S} \cite{SSL08},
\ce{^{36}Ar} \cite{SGH75}, \ce{^{40}Ca} \cite{NHZ79,MSK87} and \ce{^{44}Ti} \cite{SSK03}, 
 this paper is dedicated to the theoretical interpretation of these observables throughout light and medium-mass $\alpha$-conjugate nuclei. 
Moreover, we point out that in part of the aforementioned bibliography, e.g. the measurement of the g-factor of the $3_1^-$ state of \ce{^{16}O} in ref.~\cite{ABD84}, the 
$\alpha$-cluster nature that characterizes this nucleus was marginalized if not ignored during the discussion of the observed data. 

The primary target of the present article is to fill this gap in the literature, showing that experimental g-factor values lie in a reasonable neighbourhood of $g = +0.5$
and constitute a new unambiguous evidence for an $\alpha$-cluster structure in even-even self-conjugate nuclei. In particular, the treatment begins
with the presentation of the values of the g-factors for low-lying energy levels of light and medium mass nuclei 
computed in the framework of the nuclear shell model (cf. ref.~\cite{Bro01}). In parallel, we extract for comparison the same quantities by means of the Schmidt 
estimates \cite{Sch37,GMJ55}. Subsequently, we repeat the calculation by adding one or two neutrons to the original nuclei, 
highlighting how this operation alters the g-factor values. We further contextualize the outcoming nuclides in the realm 
of \textit{molecular nuclei} \cite{KaE19,BiI19,BiI21-02}, where the added neutron(s) either assume the role of valence particles (e.g. \ce{^{10}Be}) \cite{KaE19} 
or become an integral part of the cluster themselves (e.g. \ce{^{12}Be}) \cite{KaE19}. 

Secondly, the recent experimental advances as well as the surge of interest for $\alpha$-conjugate nuclei calls for a study of magnetic dipole moments, $\mu$, 
and g-factors in the framework of purely $\alpha$-cluster models as the one in refs.~\cite{FKK04,LLL14} on a cubic lattice. In particular, we display the behaviour 
of the average value of $\mu$ computed on the $2_1^+$ and $3_1^-$ states of \ce{^{12}C} with maximum angular momentum projection 
as a function of the lattice spacing. 
Moreover, in sec.~\ref{S-3.0} we demonstrate, how multiplet averaged \cite{LLL14} and isotropically averaged \cite{LLL15} values of the latter 
observable succeed in reducing the artifacts associated with finite lattice spacing in the two energy levels of the same nucleus
recently analysed in ref.~\cite{SEM18}.


\section{\textsf{Magnetic dipole moments}}\label{S-2.0} 

\hspace{\parindent} Magnetic dipole moments are known to contain information on the microscopic structure of a nuclear system. These 
observables are indeed very sensitive to the occupancy of quasiparticle orbits of valence nucleons and
serve as a testing ground for $A$-body wavefunctions and theoretical models in general. 
Besides, the measurement of the magnetic dipole moments represents an immediate verification 
for the assignment of a given angular momentum to nuclear ground and excited states in the shell model. 
In the many-body definition (cf. ref.~\cite{TaS63,Fio10}), the former are given by the 
superposition of the single-nucleon orbital angular momentum and spin operators \cite{Fio10},
\begin{equation}
\boldsymbol{\mu} = \frac{e}{2m_p}\sum_{\pi = 1}^Z (g_{\ell}^{p} \bfell_{\pi} + g_{s}^{p}
 \mathbf{s}_{\pi})  +  \frac{e}{2m_n}\sum_{\nu = 1}^N (g_{\ell}^{n} \bfell_{\nu} + g_{s}^{n} \mathbf{s}_{\nu})~,\label{E-2.0-01}
\end{equation}
weighted by the spin and orbital gyromagnetic factors indexed by $\pi$ ($\nu$) for protons (neutrons).  Here,
$e$ denotes the unit electromagnetic charge and $m_n$ and $m_p$ the mass of the neutron and the proton,
in order. For \textit{free} nucleons in ref.~\cite{TMN20} the g-factors assume the values
\begin{equation}
\begin{gathered}
g_{\ell}^{p} = 1 \hspace{0.5cm}g_s^{p} = 5.5856946893(16)~,~\\
g_{\ell}^{n} = 0 \hspace{0.5cm}g_s^{n} = -3.82608545(90)~.~\label{E-2.0-02}
\end{gathered}
\end{equation}
In more refined calculations, $g_s^{p}$ and $g_s^{n}$ are quenched by a factor $0.60 \div 0.75$
\cite{CaT90,PSS01} in order to account for the polarization of the core of the nucleus and for the 
meson-exchange current (MEC) corrections \cite{BDL07}. Nonetheless, some authors \cite{GRB15} prefer
to neglect these corrections and use the free values for charges and g-factors of nucleons in the microscopic
calculations of electromagnetic moments and transition probabilities. Ab-initio estimates for the corrections to the 
g-factors can be extracted from the average values of the MECs in the zero-momentum-transfer limit in 
eq.~(125) of ref.~\cite{CGP15}, derived in the framework of chiral effective field theory ($\chi$EFT). 
Note that the currents used by these authors have to be taken with a grain of salt~\cite{Kre20}.
For the Schmidt estimates (SE) presented in this section, we reduce the nucleon g-factors to $70$~\% of the free values
and assume $Z$ to be always even.


\subsection{\textsf{N = Z nuclei}}\label{S-2.1}

\hspace{\parindent}  We target the average values of the operator in eq.~\eqref{E-2.0-01} in low-energy excited states of $\alpha$-conjugate 
nuclei. Motivated by the high stability of the \ce{^4He} nucleus, we suppress the single-nucleon degrees of freedom in spite of the 
former accomplishment and interpret the nuclear excitations in terms of the rotational and vibrational motion of the $M \equiv Z/2$ $\alpha$-particles.
These bosonic groupings are characterized by zero spin and isospin as well as vanishing magnetic dipole moment in 
the $0^+$ ground state. In such a \textit{macroscopic} framework, the spin contribution to the magnetic dipole 
moment remains equal to zero in any excited state. Hence, we are allowed to set $g_{s}^{\alpha}=0$ and rewrite 
eq.~\eqref{E-2.0-01} as
\begin{equation}
\boldsymbol{\mu}^{(\alpha)} = \frac{e}{2 m_{\alpha}} \sum_{i=1}^M g_{\ell}^{\alpha}\bfell_i 
= \frac{e~g_{L}^{(\alpha)}}{2 m_p} \mathbf{L}~,~\label{E-2.1-01}
\end{equation}
where $m_\alpha$ is the $\alpha$-particle mass, the summation is performed over the $M$ $\alpha$-clusters, $g_{\ell}^{\alpha} = 2$ is the orbital component of the 
gyromagnetic factor for the $\alpha$-particle and
\begin{equation}
g_{L}^{(\alpha)} \equiv g_{\ell}^{\alpha} \frac{m_p}{m_{\alpha}} = \frac{2 m_p}{m_{\alpha}} \approx +0.5034~,~\label{E-2.1-02}
\end{equation}
coincides with the nuclear g-factor. Its small deviation with respect to $0.5$ is essentially due to the mass excess of 
the \ce{^4He} nucleus. 

\begin{figure*}[ht!]
\includegraphics[width=1.0\columnwidth]{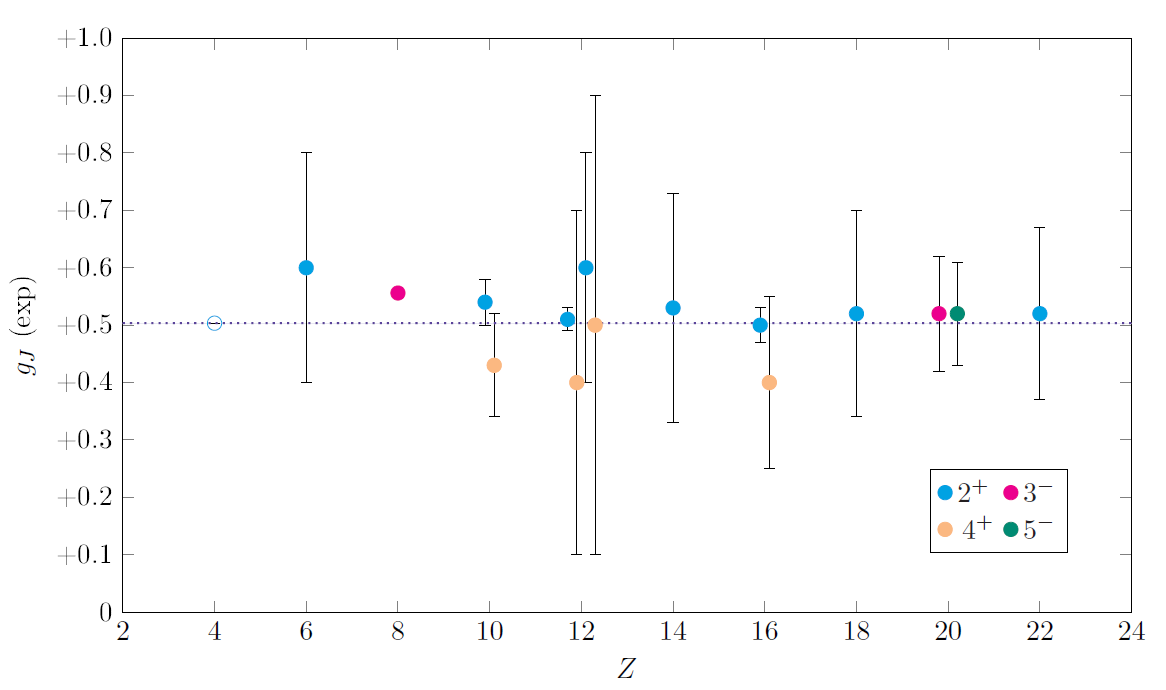}
\caption{Behaviour of the experimental g-factor values as a function of the proton number for excited states of $\alpha$-conjugate nuclei. The dotted line 
denotes the prediction from macroscopic $\alpha$-cluster models. For the nuclei \ce{^{20}Ne}, \ce{^{24}Mg}, \ce{^{32}S} and \ce{^{40}Ca} an artificial 
shift in the $Z$ direction is added in order to resolve the error bars. The theoretical $\alpha$-cluster value for the $2_1^+$ state of \ce{^8Be} is denoted 
with an open circle (cf. tab.~\ref{T-2-01})}\label{F-2-01}
\end{figure*}

The available experimental values for $g_{L}^{(\alpha)}$ in light and medium-mass $N = Z$ even-even self-conjugate nuclei
are all compatible within the statistical errors with the outcome of eq.~\eqref{E-2.1-01} for macroscopic models. 
This remarkable agreement emerges dramatically in fig.~\ref{F-2-01} and constitutes the cornerstone of the present interpretation.
The validity of this assertion is not limited to the lowest $2^+$ or $3^-$ states of $\alpha$-conjugate nuclei, but remains intact
at growing excitation energies. 

\begin{table*}[ht!]
\begin{center}
\begin{tabular}{c|c|c|ccc|c}
\toprule
\textsc{Nucleus} & \textsc{Level} & \textsc{Energy} [MeV] & $g^{(SM)}$ & $g^{(\alpha)}$ & $g^{(exp)}$ & $Z/A$\\
\midrule
\ce{^{8}Be} & $2_1^+$ & $3.030$ & $-$ & +0.5034 & $-$ & +0.50  \\
\midrule
\ce{^{12}C} & $2_1^+$ & $4.438$ & +0.507 \cite{Bro82} &  +0.5034 & +0.60(20) \cite{KHK79} & +0.50 \\
\midrule
\ce{^{16}O} & $3_1^-$ & $6.130$ & +0.555 \cite{Bro82} & +0.5034 & +0.556(4) \cite{ABD84} & +0.50 \\
\midrule 
\multirow{2}{0.65cm}{\centering{\ce{^{20}Ne}}} & $2_1^+$ & $1.634$ & +0.510 \cite{SaS17} &  \multirow{2}{1.25cm}{\centering{+0.5034}} & +0.54(4) \cite{Spe93} & \multirow{2}{0.85cm}{\centering{+0.50}} \\
& $4_1^+$ & $4.247$ & +0.513 \cite{SaS17} &   & +0.43(9) \cite{SKN02} &  \\
\midrule
\multirow{4}{0.75cm}{\centering{\ce{^{24}Mg}}} & $2_1^+$ & $1.369$  & +0.513 \cite{SaS17} & \multirow{4}{1.25cm}{\centering{+0.5034}} & +0.538(13) \cite{KSG15-01} & \multirow{4}{0.95cm}{\centering{+0.50}} \\
& $4_1^+$ & $4.123$ & +0.518 \cite{SaS17} & & +0.40(30) \cite{SMT83} & \\
& $2_2^+$ & $4.238$ & +0.519 \cite{SaS17} & & +0.60(20) \cite{SMT83} & \\
& $4_2^+$ & $6.010$ & +0.512 \cite{SaS17} & & +0.50(40) \cite{SMT84} & \\
\midrule
\ce{^{28}Si} & $2_1^+$ & $1.779$ & +0.516 \cite{SaS17} &  +0.5034 & +0.53(2) \cite{Spe93} & +0.50 \\
\midrule 
\multirow{2}{0.5cm}{\centering{\ce{^{32}S}}} & $2_1^+$ & $2.230$ & +0.505 \cite{SaS17} & \multirow{2}{1.25cm}{\centering{+0.5034}} & +0.50(3) \cite{Spe93} & \multirow{2}{0.85cm}{\centering{+0.50}} \\
& $4_1^+$ & $4.459$  &  +0.507 \cite{SaS17} &  & +0.40(15) \cite{SHS88}  &  \\
\midrule
\ce{^{36}Ar} & $2_1^+$ & $1.970$ & +0.488 \cite{SSL06} & +0.5034 & +0.52(18) \cite{SSL06} & +0.50 \\
\midrule 
\multirow{2}{0.5cm}{\centering{\ce{^{40}Ca}}} & $3_1^-$ & $3.737$ & +0.486 \cite{HTM74} & \multirow{2}{1.25cm}{\centering{+0.5034}} & +0.52(10) \cite{MSK87} & \multirow{2}{0.85cm}{\centering{+0.50}} \\
& $5_1^-$ & $4.492$  & +0.512 \cite{HTM74} & & +0.52(9) \cite{MSK87} &  \\
\midrule
\ce{^{44}Ti} & $2_1^+$ & $1.083$  & +0.514 \cite{SSK03} & +0.5034 &  +0.52(15) \cite{SSK03} & +0.50 \\
\bottomrule
\end{tabular}
\caption{Experimental nuclear g-factors for excited states of the lightest $\alpha$-conjugate nuclei (exp) with the predictions of the shell model (SM), the macroscopic $\alpha$-cluster models ($\alpha$) and the collective $Z/A$ value.}\label{T-2-01}
\end{center}
\end{table*}
Exemplary, in this respect, is the case of the $2_1^+$ and the $4_1^+$ levels of \ce{^{20}Ne}  
and \ce{^{32}S} as well as the $2_1^+$, $4_1^+$, $2_2^+$ and $4_2^+$ states of \ce{^{24}Mg}, where the measured g-factors do
 not display systematic deviations from the value in eq.~\eqref{E-2.1-01} within their sizable errors (cf. tab.~\ref{T-2-01}).
In the framework of the shell-model (SM), g-factor values in reasonable neighbourhood of the macroscopic $\alpha$-cluster value of $\approx +0.5034$ 
are found. For $10 \leq Z \leq 16$ nuclei the theoretical SM data are obtained through the phenomenological USDB interactions \cite{BrR06},
taylored for \textit{sd}-shell nuclei~\cite{Chu76,Wil84}. Conversely, for \ce{^{36}Ar} the WBT interaction in ref.~\cite{WaB92} is used in the same 
model space, whereas for \ce{^{40}Ca} the calculations \cite{HTM74} assume admixtures of random phase approximation \textit{1p-1h} states and 
deformed \textit{3p-3h} states \cite{Goo70}.
However, the SM configuration $d_{3/2}^{-1}f_{7/2}$ alone yields in \ce{^{40}Ca} g-factors in fair agreement with the experimental 
and large-scale theoretical estimates in ref.~\cite{HTM74}. Eventually, for \ce{^{44}Ti} the FPD6 \cite{RMJ91} nucleon-nucleon interaction 
is implemented in the full \textit{pf}-shell model space, assuming a self-conjugate \ce{^{40}Ca} core \cite{SSK03}.

Furthermore, for states of zero total isospin, as the bosonic ones, generated by excitations of $\alpha$-parti\-cles (cf. tab.~\ref{T-2-01}),
the g-factor is predicted to correspond to its isoscalar component,
\begin{equation}
g_J^{(0)}  = \frac{\mu_J(T_z = +T) + \mu_J(T_z = -T)}{\mu_N 2J}  = \frac{g_{\ell}^{n} + g_{\ell}^{p}}{2} + \frac{g_s^{n} + g_s^{p}
 - g_{\ell}^{n} - g_{\ell}^{p}}{2J} \langle S_z \rangle_J \label{E-2.1-03}
\end{equation} 
where $\mu_N$ is the nuclear magneton and $\mu_J(T_z = \pm T)$ are the magnetic dipole moments of two states with total angular momentum $J$ 
and opposite z-component of the total isospin, \textit{i.e.} mirror nuclei. In eq.~\eqref{E-2.1-03}, the charge dependence of nuclear force and the
 nucleon masses has been also ignored, as in eq.~(3.2) of ref.~\cite{ZHG78}. The average value of $S_z$ in the former equation is computed 
among the states with maximum total angular momentum and isospin projection along the $z$-axis, $|J, J_z = J, T, T_z = T\rangle$.

Moreover, for purely collective excitations constructed on the $J = 0$ ground state, the total angular momentum 
coincides with $\mathbf{L}$ and the average value of the third component of the spin in eq.~(3.2) of ref.~\cite{ZHG78} vanishes. 
As a consequence, the isoscalar part of the gyromagnetic factor becomes independent on $J$ and reduces to $1/2$ \cite{BBD84} (cf. eq.~\eqref{E-2.0-02}).
This result agrees with eq.~\eqref{E-2.1-01}, the deviations owing to the neglected charge dependence of nuclear forces and 
nucleon masses. 

Although the original measurements of the gyromagnetic factors of the $3_1^-$ state of \ce{^{16}O} and the $2_1^+$ state of \ce{^{24}Mg}
yielded $g=+0.55(3)$ \cite{RAC73} and $+0.51(2)$ \cite{HED75}, respectively, the more recent and more precise observations in refs.~\cite{ABD84} and \cite{KSG15-01}
highlight a $10$~\% deviation from the macroscopic $\alpha$-cluster value in eq.~\eqref{E-2.1-01}. 
This bias is only marginally covered by the shell-model calculations in ref.~\cite{WiC79}, where configuration mixing among one-body
states lying within major harmonic oscillator shells has been considered. 

Nevertheless, in ref.~\cite{Bro82} the rest of the discrepancy is successfully 
attributed to \textit{isospin mixing} exerted by the T = 1 (isovector) $3_4^-$ state at 13.26~MeV, whose reduced magnetic dipole transition 
probability to the $3_1^+$ state is large. Interpreting the two $3^-$ states as an admixture of \textit{1p-1h} neutron and proton states of an $\alpha$-particle such as
 $p_{1/2}^{-1}d_{5/2}$ and indentifying the mixing source with the Coulomb force $V_{\mathrm{C}}$, an isovector contribution with the expected 
magnitude is obtained, 
\begin{equation}
g_J^{(1)} = \frac{\mu_J(T_z = +T) - \mu_J(T_z = -T)}{\mu_N 2J} \overset{\ce{^{16}O}~3_1^-}{\approx}
 - \frac{2}{3} \frac{\langle 3^-, 1| \mu_z | 3^-, 0\rangle}{\mu_N} \frac{\langle 3^-, 0|V_{\mathrm{C}}|3^-, 1\rangle}{|\Delta E_{3^-}|}~,\label{E-2.1-04}
\end{equation}
where $|\Delta E_{3^-}|$ denotes the energy separation between the two levels and the compact notation $|J, T \rangle \equiv |J T J_z T_z\rangle$
for the shell-model states is understood. In summary, the full gyromagnetic factor of the $3_1^-$ state, $g_J^{(0)} + g_J^{(1)}$, becomes $+0.555$ \cite{Bro82}. 
The same considerations, applied to the  $2_1^+$ (T = 0) state at $4.44$~MeV and the $2_4^+$ (T = 1) state at $16.11$~MeV of \ce{^{12}C}, result in a
smaller isovector correction to the g-factor of about $1.5$~\% with respect to the shell-model value (cf. tab.~\ref{T-2-01}). Altogether, the main
nuclear structure lends weight to the consistency of $\alpha$-clustering.

Concerning the $2_1^+$ state of \ce{^{24}Mg}, the gyromagnetic factor from the USDB Hamlitonian in ref.~\cite{BrR06}, together with
the isovector correction (cf. ref.~\cite{OrB88}) yields $+0.520$, is consistent with the most recent experimental
counterpart \cite{KSG15-02}. The overlap improves when also the effects of the MECs on the bare nucleon g-factors in eq.~\eqref{E-2.0-02}
are included \cite{ASB87}, leading to a final value of $+0.544$ \cite{KSG15-02}.
For the other values of the gyromagnetic factors reported in tab.~\ref{T-2-01}, times are not ripe for singling out
possible deviations from the macroscopic $\alpha$-cluster value in eq.~\eqref{E-2.1-01}, since the statistical errors affecting the
measurements overcome on average the $10$~\% of the g-factor value itself.

Despite the sizable experimental errors affecting the available estimates and the small deviations from $g_J^{(0)}$, our $\alpha$-cluster 
interpretation of the g-factors for this class of nuclei remains solid.
Underpinning this construction is the gradual emergence of $\alpha$-clu\-stering at increasing excitation energy and, remarkably, close
to $\alpha$-decay thresholds \cite{ITH68}. 
Although yet unmeasured, the magnetic moments of the $3_1^-$ and 
the $2_2^+$ states of \ce{^{12}C} as well as the $3_5^-$ and $5_1^+$ states of \ce{^{16}O} in the close vicinity
of the $3\alpha$- and $4\alpha$-decay thresholds, respectively, are expected to adhere even more faithfully 
with the predictions  of the macroscopic $\alpha$-cluster picture. 

Besides, as shown in sec.~\ref{S-2.2}, the observed independence of the g-factors on the angular momentum 
and the excitation energy of the states to which they refer, is a peculiar feature, which the $N = Z+1$ and $Z+2$ 
nuclei with even $Z$ do not preserve.\\


\subsection{\textsf{N = Z + 1 nuclei}}\label{S-2.2}

\hspace{\parindent} In the following we consider the case in which a single neutron is added to an $\alpha$-conjugate nucleus. The major question that naturally arises is 
whether the additional neutron enhances or weakens $\alpha$-clustering. Although no general
answer exists \cite{KaE19}, clustered states are expected to appear in correspondence of the $M\alpha + n$ decay thresholds.
For \ce{^9Be}, different approaches such as the linear combination of atomic 
orbitals (LCAO) \cite{OAT77} and the cluster shell model (CSM) \cite{DRI18} highlight an underlying dumbbell configuration with the added
neutron moving in the cluster field with $Z_2$ symmetry \cite{DRI18}. 

Concerning \ce{^{13}C}, the states with excitation energy lying between $9.90$ and $16.95$~MeV are well reproduced by a molecular 
$\alpha$-cluster model \cite{MiO02}, in which the odd neutron lies between the \ce{^{4}He} nuclei arranged in a linear chain. 
Besides, for the $(1/2)_2^-$ and $(1/2)_1^+$ states of the same nucleus, the triangular $\alpha$-cluster configuration with 
one neutron located in the same plane of the \ce{^4He} nuclei ($\pi$-orbit) and along the axis perpendicular to the $3\alpha$-plane
($\sigma$-orbit), respectively, has been proposed \cite{MiO02}.
Additionally, the large isoscalar reduced electric monopole transition probability between the $(1/2)_2 ^-$ and $(1/2)_3 ^-$ states
and the ground state \cite{SKU06} is also regarded as a signature of $\alpha$-clustering in this nucleus \cite{BiI19}.
Indeed, states lying below the \ce{^9Be}+$\alpha$ decay thres\-hold are well predicted by the shell-model for \textit{p}-shell nuclei \cite{CoK65}. 

\begin{table}[ht!]
\begin{center}
\begin{tabular}{c|c|c|ccc|c}
\toprule
\textsc{Nucleus} & \textsc{Level} & \textsc{Energy} [MeV] & $g^{(SM)}$  & $g^{(SE)}$ & $g^{(exp)}$ & $Z/A$ \\
\midrule
\ce{^{9}Be} & $\frac{3}{2}_1^-$ & $0.0$ & -0.713 \cite{HWG88} & -0.89 & -0.784955(2) \cite{Ita83} & +0.444\\
\midrule
\multirow{2}{0.75cm}{\centering{\ce{^{13}C}}} & $\frac{1}{2}_1^-$ & $0.0$  & +1.506 \cite{HWG88} & +0.89 & +1.404824(2) \cite{Roy54} & +0.462 \\
 & $\frac{5}{2}_1^+$ & $3.854$ & -0.597 \cite{HWG88}  & -0.77 & -0.59(5) \cite{BAA74} & +0.462\\
\midrule
\ce{^{17}O} & $\frac{5}{2}_1^+$ & $0.0$ & -0.7652 \cite{SaS17} & -0.54 &  -0.75752(4) \cite{AlY51-01} & +0.471\\
\midrule
\multirow{2}{0.75cm}{\centering{\ce{^{21}Ne}}} & $\frac{3}{2}_1^+$ & $0.0$ & -0.500 \cite{SaS17} & $-$ & -0.441198(3) \cite{BBC77} & +0.476\\
& $\frac{5}{2}_1^+$ & 0.351 & -0.230 \cite{SaS17} & $-0.54$ & -0.196(12) \cite{RAD78} & +0.476\\
\midrule
\ce{^{25}Mg} & $\frac{5}{2}_1^+$ & $0.0$  & -0.340 \cite{SaS17} & -0.54 & -0.34218(3) \cite{AlY51-02} & +0.480 \\
\midrule
\ce{^{29}Si} & $\frac{1}{2}_1^+$ & $0.0$  & -1.114 \cite{SaS17} &  -2.67 & -1.1106(6) \cite{Wea53} & +0.483 \\
\midrule
\ce{^{33}S} & $\frac{3}{2}_1^+$ & $0.0$  & +0.387 \cite{MGW80} & +0.54  & +0.4292141(9) \cite{LNS73} & +0.485 \\
\midrule
\multirow{2}{0.75cm}{\centering{\ce{^{37}Ar}}} & $\frac{3}{2}_1^+$ & $0.0$  & +0.601 \cite{WiB65} &  +0.54 & +0.763(3) \cite{BAPS88} & +0.486 \\
& $\frac{7}{2}_1^-$ & $1.611$  & -0.43 \cite{MCJ75} &  -0.38 & -0.38(2) \cite{RBM71} & +0.486 \\
\midrule
\ce{^{41}Ca} & $\frac{7}{2}_1^-$ & $0.0$  & -0.49 \cite{GRB15} & -0.38 & -0.455652(3) \cite{BKP62}  & +0.488 \\
\midrule
\multirow{2}{0.75cm}{\centering{\ce{^{45}Ti}}} & $\frac{7}{2}_1^-$ & $0.0$  & +0.159 \cite{KBO78} & -0.38  & +0.027(1) \cite{CoM66} & +0.489\\
& $\frac{5}{2}_1^-$ & $0.0393$  & -0.292 \cite{KBO78} & +0.38 & -0.053(4) \cite{BRA77} & +0.489\\
\bottomrule
\end{tabular}
\caption{Nuclear g-factors for the ground and excited states of the lightest $N = Z + 1$ nuclei, replacing the $\alpha$-cluster values in tab.~\ref{T-2-01} by the Schmidt estimates (SE). The negative sign of the reported values reflects convincingly the negative magnetic moment of the added neutron. In contrast with tab.~\ref{T-2-01}, the ratios $Z/A$ do not agree with the experimental g-factors.}\label{T-2-03}
\end{center}
\end{table}
Concerning \ce{^{17}O}, attempts to reproduce the spectrum of the nucleus above the \ce{^{13}C}+$\alpha$ decay threshold 
in terms of the $4\alpha +n$ configuration have been made \cite{DuD01,DuD05} in the framework of the generator coordinate 
method (GCM) \cite{Bri66}. Nevertheless, for the lowest-energy states, SM-based approaches as the cluster-orbital shell model 
(COSM) \cite{MKI06} seem to provide a rather faithful description \cite{KSF08,MOM09}.
Moreover, a recent CSM analysis of the whole low-energy spectrum and the electromagnetic multipole transition probabilities
 of \ce{^{21}Ne} \cite{BiI21-02} suggests, that particle and hole neutron states can coexist with the underlying \ce{^{20}Ne} core, thus 
exciting the internal degrees of freedom of one of the $\alpha$-particles. The comparison with the experimental energies turns out 
to be favourable for the model, except for the missing $K^P = 1/2^+$ band starting at $17.34$~MeV \cite{BiI21-02}. Besides, the g-factor
of the $(3/2)_1^+$ state, inferred from the magnetic moment \cite{BiI21-02} with the free value for $g_s^{n}$, 
is equal to $-0.287$, \textit{i.e.} smaller in magntitude than the experimental one in tab.~\ref{T-2-03}. The value is obtained by
addition of the g-factor of the $5\alpha$ bi-piramidal core (cf. eq.~\eqref{E-2.1-01}) with the contribution of the unpaired neutron,
$ -0.85$~$\mu_N$, times suitable Clebsch-Gordan coefficients (cf. eq.~(20) in ref.~\cite{BiI21-02}).

Switching to heavier nuclei, the cross section of elastic and inelastic scattering of $\alpha$-particles suggests that 
$\alpha$-particles at least at the surface level are present and interact with the incoming $\alpha$-particle \cite{SzT76}. Moreover, in \ce{^{41}Ca} and 
\ce{^{45}Ti}, $\alpha$-clusters can be induced by the pairing interaction alone at large values of the strength parameter $G$ or at physical
values of the latter, provi\-ded a four-nucleon force is introduced in the Hamiltonian \cite{SzT76}. 

\begin{figure}[ht!]
\includegraphics[width=1.0\columnwidth]{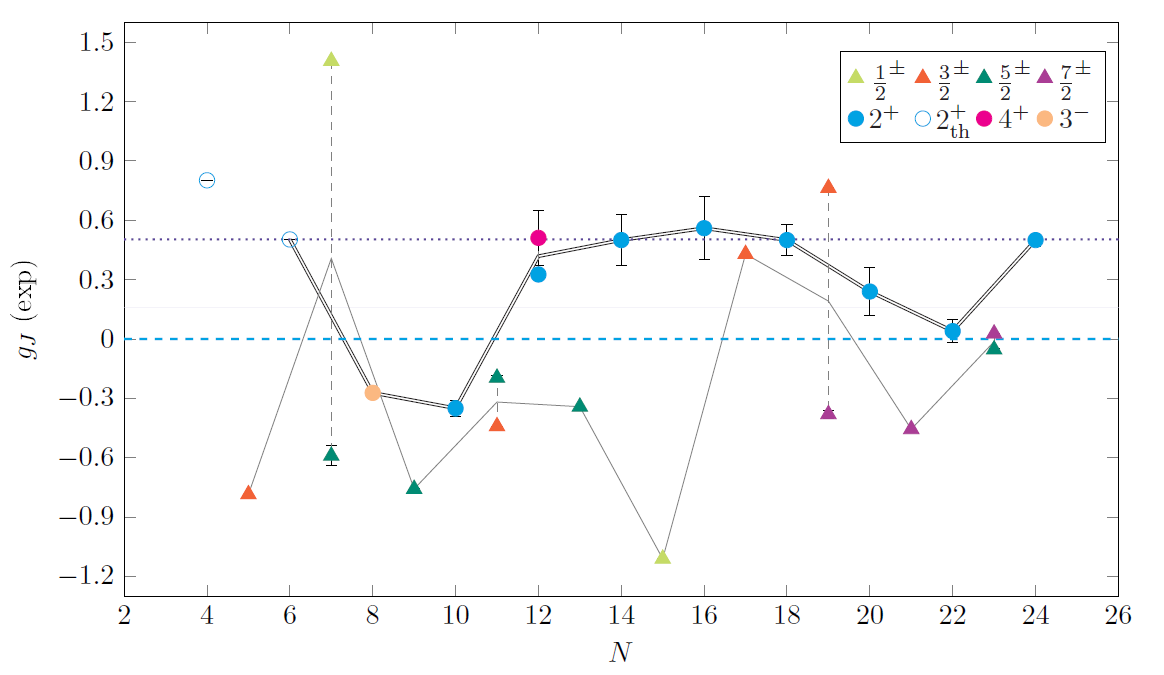}
\caption{Behaviour of the experimental g-factor values as a function of the neutron number for ground and excited states of odd nuclei with $N = Z + 1$ (triangles) and even-even nuclei with $N = Z + 2$ (circles). The theoretical values available for \ce{^{10}Be} and \ce{^{10}C} are denoted with an open circle. The double (single) line joins the averaged values of the g-factors for the ground and excited states (when present) related to the $N = Z + 2$ ($N = Z + 1$) nuclei.}\label{F-2-02}
\end{figure}

The experimental data of the gyromagnetic fac\-tors for this class of nuclei (cf. tab.~\ref{T-2-03}) highlight large deviations from
the $\alpha$-cluster value in eq.~\eqref{E-2.1-02}, as a result of the significant spin contribution from the unpaired neutron. 
The fact itself that the magnetic moment is not parallel to the total angular momentum leads to strongly J-dependent values 
of the g-factor in excited states of the same nucleus. Exemplary is the case of the $(1/2)_1^-$ and the $(5/2)_1^+$ states of \ce{^{13}C}
(cf. fig.~\ref{F-2-02}), whose g-factors differ by almost two units and the sign.

As a consequence, we replace the macroscopic $\alpha$-cluster result in eq.~\eqref{E-2.1-01} by the Schmidt formula \cite{Sch37}, 
that attaches the magnetic moment to the unpaired nucleon, 
\begin{equation}
\boldsymbol{\mu} = \frac{e}{2m_u} (g_{\ell}^{u} \bfell + g_{s}^{u} \mathbf{s})~,\label{E-2.2-01}
\end{equation}
where $u = n$ or $p$. Due to the inequality between the orbital and the spin gyromagnetic factors, $\boldsymbol{\mu}$ precedes 
about the angular momentum $\mathbf{j} = \bfell + \mathbf{s}$ of the single-particle ground or excited state. The 
reported values of the nuclear g-factors are extracted from the magnetic moments by taking the projection of 
$\boldsymbol{\mu}$ 
on the total angular momentum operator of the nucleus, $\mathbf{J}$,
\begin{equation}
\boldsymbol{\mu}_J \equiv \frac{e}{2m_u}  g_J \mathbf{J} = \frac{\mathbf{J} 
\cdot (g_{\ell}^u \bfell + g_s^u \mathbf{s}) }{J(J+1) \hbar^2}\frac{e\mathbf{J}}{2m_u}~,\label{E-2.2-03}
\end{equation}
where $\mathbf{J}$ is assumed to coincide with the one of the single-particle 
state $\mathbf{j}$ and the inner products are evaluated via the Landé formulae,
\begin{equation}
\bfell \cdot \mathbf{J} = \frac{\hbar^2}{2} \left[J(J+1) + \ell(\ell+1) - s(s+1)\right]~,\label{E-2.2-04}
\end{equation}
\begin{equation}
\mathbf{s} \cdot \mathbf{J} = \frac{\hbar^2}{2} \left[J(J+1) + s(s+1) - \ell(\ell+1)\right]~.\label{E-2.2-05}
\end{equation}
The description of $\boldsymbol{\mu}$, merely in terms of the unpaired particle, proved to be quite effective for nuclei in 
the vicinity of a closed shell \cite{RiS04}. In addition, the adoption of quenched values for $g_s^{p}$ and $g_s^{n}$ 
in the Schmidt estimates for the gyromagnetic factors permits to improve the estimates for the open-shell nuclei, at the price
of deteriorating slightly the good agreement with the experimental values of the \ce{^{17}O} and \ce{^{41}Ca} nuclei
(cf. tab.~\ref{T-2-03} and fig.~\ref{F-2-02}).

In fact, the Schmidt estimates for the \emph{sd}-shell nuclei improve and quite accurate predictions are  
obtained for the $(5/2)_1^+$ state of \ce{^{25}Mg}, as well as for the $(3/2)_1^+$ one of \ce{^{33}S}
and the $(3/2)_1^+$ and $(7/2)_1^-$ of \ce{^{37}Ar}. Conversely, the values of $g^{(SE)}$ associated with the $(7/2)_1^-$ ground state and the $(5/2)_1^-$ excited state 
of \ce{^{45}Ti}, sharply disagree with the experimental data. It follows that a larger model space is required for this nucleus, lying in the 
\emph{pf}-shell, in order to obtain order-of-magnitude agreement with the observed g-factors. 
The small excitation energy of the $(5/2)_1^-$ state itself is an indicator of \textit{collective motion}.

Concerning the shell-model estimates for \ce{^{9}Be}, calculations with phenomenological interactions encompassing the active 
nucleons in the \textit{p}-shell ($0~\hbar\omega$ space) as well as the excitations to the \textit{sd}-shell ($1~\hbar\omega$ space) produce a
g-factor for the $(3/2)_1^-$ state differing by less than 10\% from the experimental counterpart \cite{HWG88}. The deviation 
reduces to nearly $1$~\% in the microscopic $2\alpha+n$ model in ref.~\cite{AOS96}, where $g \approx -0.779$, thus lending further weight to
the \textit{molecular} treatment of the unpaired nucleon moving around the $\alpha$-clusters.

For \ce{^{13}C}, the former shell-model calculation with phenomenological interaction yields a similar agreement with the experimental
g-factor of the ground state, whereas a more recent one based on Warburton-Brown interactions underestimates the same 
g-factor by $10$~\% \cite{YSO12}. The closest available estimate to the measured g-factor in tab.~\ref{T-2-03} yields $+0.707$ \cite{SFO03}
and is based on the \ce{^4He} core in the \textit{p-sd} model space and with $2-3~\hbar\omega$
excitations, adopting the bare values of the proton and neutron g-factors. 
The result for the $(5/2)_1^+$ excited state turns out to be even closer to the experimental value, which
is indeed well reproduced ($g \approx -0.62$) by the minimal coupling model in ref.~\cite{MeP75}, that assumes a self-conjugate \ce{^{12}C} core.

On the other hand, the g-factors of the \textit{sd}-shell nuclei up to \ce{^{29}Si} are obtained from USDB interaction \cite{BrR06}, 
and some disagreement with the experimental data is detected only for the $(3/2)_1 ^+$ and $(5/2)_1 ^+$ states of 
\ce{^{21}Ne} \cite{SaS17}. The discrepancy is slightly filled by the ab-initio in-medium similarity renormalization group (IM-SRG) approach,
yielding $g = -0.443$ and $-0.140$ for the two states, respectively.

Concerning the $(3/2)_1^+$ of \ce{^{33}S}, the \textit{sd}-shell model calculation in ref.~\cite{MGW80} with the empirical Hamiltonian 
of ref.~\cite{Chu76} reproduces the experimental g-factor within less than $10$~\% deviation. Moreover, the gyromagnetic factor of the Schmidt estimate 
approaches the measured value, thus suggesting that the largest contribution comes from the odd neutron in the $d_{3/2}$ shell \cite{MGW80}.

Assuming a \ce{^{28}Si} core with configuration mixing and effective $g_s^{p}$ and $g_s^{n}$ for the $1s_{1/2}$ and $0d_{3/2}$ shells, 
the g-factor of the analogous $(3/2)_1^+$ state of \ce{^{37}Ar} has been computed in ref.~\cite{WiB65}, when the experimental counterpart was 
unavailable. From the different shell-model estimates for the magnetic dipole moments of tab.~1 in ref.~\cite{WiB65} follows, that a better 
agreement with the observed value of this state is obtained when the g-factors of the free nucleons are adopted \cite{WiB65}. For the $(7/2)_1^+$
state of the same nucleus, in tab.~\ref{T-2-03} the shell-model state with negative parity is constructed from a 
\ce{^{36}Ar} core with a single neutron in the \textit{pf}-shell, exploiting the free nucleon g-factors \cite{MCJ75}. Core polarization effects
are included in the calculation, by considering both the $0_1^+$ and the $1^+$ levels of the $\alpha$-conjugate core. The resulting g-factor
is compatible within four standard deviations with the experimental value.

In the \ce{^{41}Ca} case, the reference result arises from a large scale \textit{sd-pf} shell-model calculation with \ce{^{28}Si} core
and USD \cite{Wil84}, and modified Kuo-Brown \cite{PoZ81} interactions for the \textit{sd} and \textit{pf} shells, respectively, and LKS \cite{KLS69} interaction 
for the mixing of the two major shells. The g-factor in tab.~\ref{T-2-03} is found to reproduce with less than 10\% deviation the experimental 
value, even if the free values for $g_s^{p}$ and $g_s^{n}$ are employed.

Finally, the gyromagnetic factors of the $(7/2)_1^-$ and $(5/2)_1^-$ states of \ce{^{45}Ti} are drawn from the shell model calculation 
in the \textit{pf} model space with phenomenological interactions in ref.~\cite{KBO78}, in which $g_s^{p}$ and $g_s^{n}$ in the \textit{pf} shell
are fitted to reproduce the experimental magnetic moment of the ground state of \ce{^{41}Ca} (and \ce{^{41}Sc}). The g-factors of the two levels of \ce{^{45}Ti}
in tab.~\ref{T-2-03} turn out to be overestimated in magnitude, although less than the Schmidt estimates, but possess the correct sign. Nevertheless, the application of the Nilsson model
in ref.~\cite{Law61} with oblate deformation yields a g-factor of $-0.003$ for the $7/2^-$ ground state \cite{Law61}, underestimated 
but sensibly closer to the measured counterpart.\\


\subsection{\textsf{N = Z + 2 nuclei}}\label{S-2.3}

\hspace{\parindent} In the following, we add another neutron to the original even-even self-conjugate nuclei and investigate, how the two extra neutrons affect the
inner $\alpha$-cluster structure. 

As in \ce{^9Be} case, for the \ce{^{10}Be} nucleus, a number of studies examining the low-lying spectrum 
and the reduced electric and magnetic transition probabilities exists in microscopic or semi-microscopic pictures.
Among these, the LCAO \cite{ItO00} and the antisymmetrized molecular dynamics (AMD) \cite{KEH99} have established that $\alpha$-clustering in \ce{^{10}Be}
emerges already in the ground and excited states of the $K^P = 0^+$ band, where the two valence neutrons form structures equivalent to $\pi$ bonds. The
separation between the \ce{^4He} clusters grows in the second $K^P = 0^+$ and in the negative parity band $K^P = 1^-$, where the two neutrons give 
rise to $\sigma$-like molecular bonds \cite{KEH99}. 

\begin{table}[h!]
\begin{center}
\begin{tabular}{c|c|c|ccc|c}
\toprule
\textsc{Nucleus} & \textsc{Level} & \textsc{Energy} [MeV] & $g^{(SM)}$  & $g^{(SE)}$ & $g^{(exp)}$ & $Z/A$ \\
\midrule
\ce{^{10}Be} & $2_1^+$ & $3.368$  & +0.503 \cite{OtsYY,LiL14}  & -0.45 & $-$ & +0.400 \\
\midrule
\ce{^{10}C} & $2_1^+$ & $3.354$  & +0.802 \cite{OtsYY} & +1.99 & $-$  & +0.600 \\
\midrule
\ce{^{14}C} & $3_1^-$ & $6.728$ & -0.261 \cite{ACF74} & -0.30 & -0.272(7) \cite{ACF74} & +0.429\\
\midrule
\ce{^{18}O} & $2_1^+$ & $1.982$ & -0.3995 \cite{SaS17} & -0.20 & -0.35(4) \cite{SGH75} & +0.444 \\
\midrule
\multirow{2}{0.75cm}{\centering{\ce{^{22}Ne}}} & $2_1^+$ & $1.982$ & +0.374 \cite{SaS17} & -0.19 & +0.326(12) \cite{BBD84} & \multirow{2}{0.85cm}{\centering{+0.455}} \\
& $4_1^+$ & $3.357$ & +0.511 \cite{SaS17} & -0.14  & +0.55(14) \cite{BBD84} & \\
\midrule
\ce{^{26}Mg} & $2_1^+$ & $1.809$ & +0.8695 \cite{SaS17} & -0.19 & +0.50(13) \cite{SKK81} & +0.462 \\
\midrule
\ce{^{30}Si} & $2_1^+$ & $2.235$ & +0.366 \cite{SaS17} & -0.27  & +0.56(16) \cite{EHD75} & +0.467 \\
\midrule
\ce{^{34}S} & $2_1^+$ & $2.128$ & +0.50 \cite{WiC79} & -0.27 & +0.50(8) \cite{ZHR79} & +0.471 \\
\midrule
\ce{^{38}Ar} & $2_1^+$ & $2.167$ & +0.309 \cite{SSL06} & -0.84 & +0.24(12) \cite{SSL06} & +0.474 \\
\midrule
\ce{^{42}Ca} & $2_1^+$ & $1.525$ & +0.13 \cite{SHS03} & -0.13 & +0.04(6) \cite{SHS03} & +0.476 \\
\midrule
\ce{^{46}Ti} & $2_1^+$ & $0.889$ & +0.285 \cite{ESK00} & -0.13 & +0.50(3) \cite{ESK00} & +0.478  \\
\bottomrule
\end{tabular}
\caption{Nuclear g-factors for excited states of $N = Z + 2$ nuclei, including \ce{^{10}C}. As noticed in tab.~\ref{T-2-03}, the ratios $Z/A$ do not follow with the measured g-factors.}\label{T-2-05}
\end{center}
\end{table}
Moreover, the application of the AMD in combination of the GCM in ref.~\cite{SuK10} unveiled that the $0_1^+$, the $2_1^+$ and the $3_1^-$ states of \ce{^{14}C} possess 
intermediate features between a triaxial $\alpha$-clustered configuration and a SM state, in which the two neutrons lie in the closed \textit{p}-shell, while
the protons fill the $p_{3/2}$ level. 
On the other hand, the $0_2^+$, $2_2^+$ and $4_1^+$ have a significant overlap with a triangular $\alpha$-cluster
configuration, in which the two neutrons stand in the same plane of the \ce{^4He} clusters and fill two \textit{sd}-like orbitals \cite{SuK10}. Linear-chain
$\alpha$-cluster states are, indeed, found to characterize the $0_5^+$, $2_6^+$ and $4_6^+$ states located in the vicinity of the $\alpha$ + \ce{^{10}Be} 
decay threshold \cite{SuK10}. 

Concerning \ce{^{18}O}, the levels $0_1^+$, $2_1^+$ and $4_1^+$ below the \ce{^{14}C}+$\alpha$ threshold in the GCM model in ref.~\cite{FKD08} 
have a strong overlap with the SM $0~\hbar\omega$ prolate or spherical configuration, whereas the states belonging to the second $K^P = 0^+$
band support the formation of the inner $\alpha$+\ce{^{14}C} structure \cite{FKD08}. 

A similar study \cite{DuD03} conducted on \ce{^{22}Ne} predicts the existence of a $K^P = 0^+$ $\alpha$-clustered band as well as negative parity doublets located 
above the \ce{^{18}O}+$\alpha$ threshold. Although less abundant, some theoretical cluster studies for the heavier $\alpha$-conjugate nuclei are available,
such as the ones in the framework of the weak coupling approach of the $\alpha$-particle (or hole) with respect to a \ce{^{40}Ca} core \cite{MiO94}, or in the 
semi-microscopic algebraic cluster model (SACM) \cite{Lev13,FLS96}. The existence of parity doublet bands with $K^P = 0^+$ and $0^-$ in the neighbourhood
of the lowest $\alpha$-decay threshold is considered as a signature for the decomposition of the nucleus into a core plus an $\alpha$-particle \cite{MiO94}. 

Regarding the g-factors of excited energy levels, their values are evidently correlated to the shell-closures. In particular, from the behaviour of 
the double lines in fig.~\ref{F-2-02}, we infer that in the semi-magic nuclei at the $N=8$ (resp. $20$) shell closures, namely \ce{^{14}C} and 
\ce{^{18}O} (resp. \ce{^{38}Ar} and \ce{^{42}Ca}) the effect of the two extra neutrons is maximum (cf. tab.~\ref{T-2-05}). For these nuclei, single-particle estimates are 
expected to reproduce quite accurately the g-factors, even when the free values of $g_s^{p}$ and $g_s^{n}$ are adopted. 

\begin{figure}[ht!]
\includegraphics[width=1.0\columnwidth]{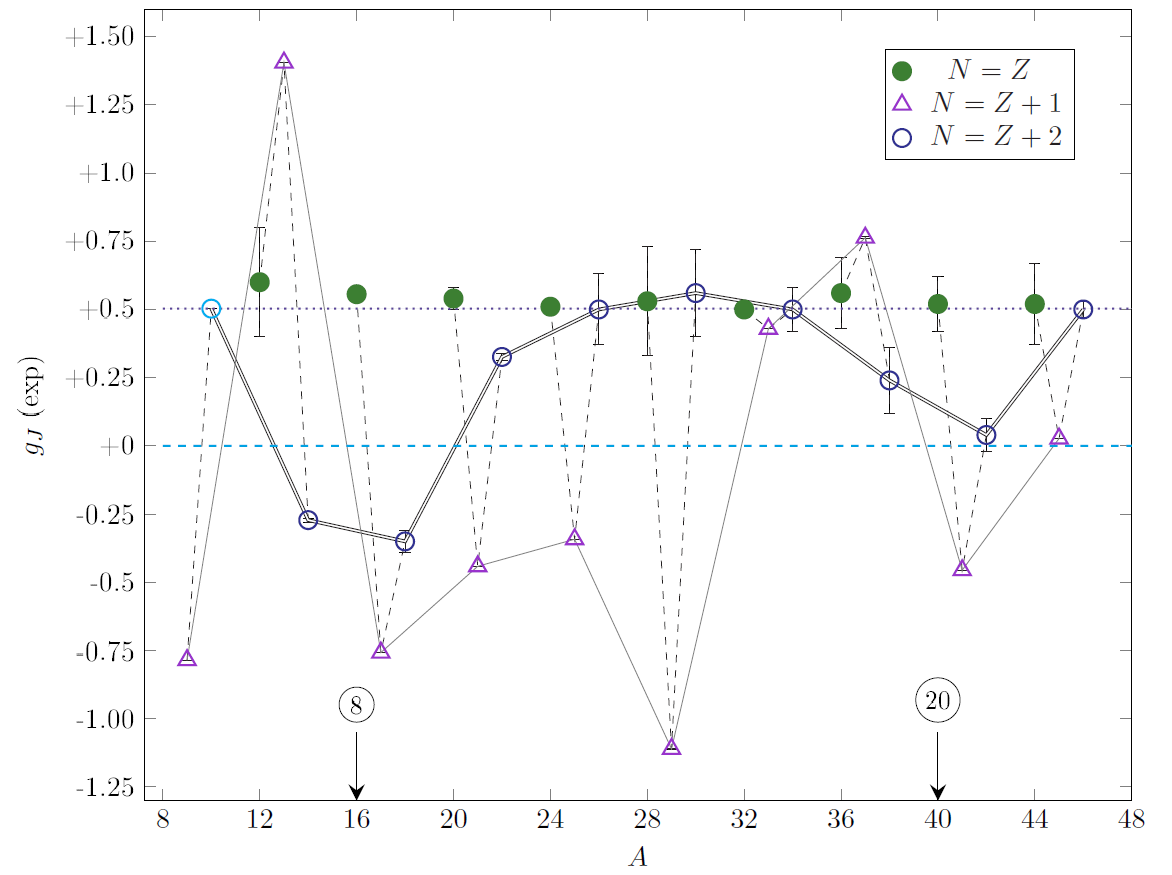}
\caption{Behaviour of the experimental g-factor values as a function of the mass number for ground and/or excited states of $\alpha$-conjugate nuclei (full circles), 
odd nuclei with $N = Z + 1$ (open triangles) and even-even nuclei with $N = Z + 2$ (open circles). The dotted line
at $g_J \approx 0.5034$ denotes the gyromagnetic factor predicted by macroscopic $\alpha$-cluster models. The theoretical value available for \ce{^{10}Be} is depicted in a lighter shade.
The shell closures at $N,Z = 8$ and $20$ are marked by arrows. Gyromagnetic factors referring to isotopes of the same nucleus are connected by dashed lines. 
The double (single) lines join the averaged values of the g-factors for the ground and excited states (when present) related to the $N = Z + 2$ ($N = Z + 1$) nuclei.}\label{F-2-03}
\end{figure}
For the other nuclei, located before, between or beyond the two shell closures (cf. tab.~\ref{T-2-05}), no significant discrepancy is found with respect to the
macroscopic $\alpha$-cluster prediction, with the sole exception of the g-factor of the $2_1^+$ state of \ce{^{22}Ne}. For all other nuclei, the experimental 
g-factors agree within one standard deviation with $\approx +0.5034$. Additionally, for the $4_1^+$ state of \ce{^{22}Ne}, the g-factor turns out to be again compatible 
with the prediction of eq.~\eqref{E-2.1-02}. Since the pairing force couples equal nucleons with opposite total angular momentum
projection, this interaction is not responsible for the suppression of the isovector contribution to the g-factor. On the other hand, four-body correlations between pairs of equal nucleons can 
quench $g_J^{(1)}$ and reduce the $g_J^{(0)}$ to $1/2$, due to the vanishing total spin. 

The application of AMD~+~GCM in ref.~\cite{SuK10} to the low-lying states of \ce{^{14}C},
suggests that the considered $3_1^-$ level in tab.~\ref{T-2-03} has a spatial distribution, intermediate between a SM state with the protons in the $p_{3/2}$ subshell
and the neutrons in the\textit{1p-1h} excited configuration and a triangular $\alpha$-cluster state. In fact, the single-particle Gaussian wave-packets of the two neutrons gather 
still around the origin \cite{SuK10}, despite a slight overall triaxial deformation ($\beta \approx 0.22$, $\gamma \approx 24^{\circ}$). Also, the GCM analysis in ref.~\cite{FKD08}
predicts a strong overlap with a SM state for the $2_1^+$ state of \ce{^{18}O} in tab.~\ref{T-2-05}. We can deduce that,
when the wavefunctions of the two extra neutrons overlap significantly in the space, clustering is hindered \cite{SSM13} and the g-factor displays significant deviations from eq.~\eqref{E-2.1-02}. 
Otherwise, if the two neutrons are spatially more separated, \textit{e.g.} when sitting between two different pairs of $\alpha$-particles in the molecular picture \cite{SSM13}, the g-factor
approaches the value of $\approx 0.5034$, as in $\alpha$-conjugate nuclei.

Therefore, we can predict that at higher-lying energy levels of \ce{^{14}C} and \ce{^{18}O} such as the $2_2^+$ state, where clustering is expected to be more pronounced 
\cite{FKD08,SuK10}, the g-factor reaches the one of the other levels of the open-shell even-even $N = Z+2$ nuclei in tab.~\ref{T-2-05}. The shell closures at $N =8$ and $20$ (cf. fig.~\ref{F-2-03}), in fact, 
seem to favour a more compact spatial distribution of the two neutrons, that undermines $\alpha$-clustering in the low-lying energy states of the nearby semi-magic nuclei, 
(cf. \ce{^{14}C}, \ce{^{18}O}, \ce{^{38}Ar} and \ce{^{42}Ca} in tab.~\ref{T-2-05}). 

Although of a milder type, the phenomenon takes place also in the two considered states
of \ce{^{22}Ne}, where the measured g-factor of the $2_1^+$ state differs by 35~\% with respect to the macroscopic $\alpha$-clu\-ster value.\\
As in the previous section, we give  the shell-model g-factors in tab.~\ref{T-2-03} and the Schmidt estimates, that we obtain through the quenched values of $g_s^{p}$ and $g_s^{n}$. The g-factor of the excited states of this class of nuclei is now attributed to two neutrons lying on two distinct single-particle states with angular momentum 
$\mathbf{j}_1$ and $\mathbf{j}_2$, which sums up to $\mathbf{J}$. 

Next, the g-factors corresponding to the two single-particle levels are combined together, giving the overall
$g_J$ factor through the formula
\begin{equation}
g_J = \frac{g_1 \mathbf{j}_1 + g_2 \mathbf{j}_2}{J(J+1)\hbar^2}\cdot \mathbf{J}  = \frac{1}{2}(g_1 + g_2) 
 + \frac{1}{2}\frac{j_1(j_1 + 1) - j_2(j_2 + 1)}{J(J+1)}(g_1 - g_2)~.\label{E-2.3-01}
\end{equation}
Applying eq.~\eqref{E-2.3-01} we obtain the values of the g-factors reported in the $g^{(SE)}$ column of tab.~\ref{T-2-05}. 
As it can be observed, the Schmidt estimates for the listed levels of \ce{^{14}C}, \ce{^{18}O} and \ce{^{42}Ca} at the chosen value 
of $g_s^{n}$ prove to be quite predictive, whereas for the $2_1^+$ state of \ce{^{22}Ne}, \ce{^{26}Mg}, \ce{^{30}Si}, \ce{^{34}S}, \ce{^{38}Ar} and 
\ce{^{46}Ti} large discrepancies, including the sign, are found with the experimental values. This fact highlights the sensitivity of the magnetic dipole 
moment to shell-closure effects in semi-magic nuclei as \ce{^{14}C}, \ce{^{18}O} and \ce{^{42}Ca}, where the g-factors are determined by either the two 
extra neutrons or the two missing protons.

Concerning the SM estimates, the gyromagnetic factor of the $2_1^+$ state of \ce{^{10}Be}, calculated via the Monte Carlo shell-model (MCSM) with
nucleon-nu\-cleon interactions  drawn from Chiral Effective Field Theory in the unitary correlation operator method (UCOM) \cite{FNR98,RRH08}, turns out to reproduce faithfully the value of eq.~\eqref{E-2.1-02}, thus suggesting a well-developed
$\alpha$-cluster structure, with the two neutrons acting as valence particles in a $\pi$ orbital of a binary molecule (cf. ref.~\cite{KEH99}). Although \ce{^{10}Be}
is a rather long-lived isotope, the magnetic moments associated with its excited states have not yet been measured. 

For \ce{^{14}C}, the SM calculation, reported in ref.~\cite{ACF74}, is based on a \ce{^{16}O} core with three holes in the \textit{p}-shell and one particle in the
\textit{sd}-shell. Besides, for the nucleons in the \textit{p}-shell, the Cohen-Kurath two-body matrix elements \cite{CoK65} are adopted, whereas the Gillet interaction 
\cite{GVM64} is exploited between the \textit{p} and \textit{sd}-shell nucleons. The result is closer in magnitude with respect to the more recent estimate in 
ref.~\cite{YSO12}, obtained with the phenomenological $V_{MU}$, SFO and SDPF-M interactions. 

The data obtained from the phenomenological USDB interaction \cite{BrR06} for the \textit{sd}-shell (cf. tab.~I in ref.~\cite{SaS17}) permit to reproduce quite
accurately the g-factors of the $2_1^+$ state of \ce{^{18}O} as well as the $2_1^+$ and $4_1^+$ states of \ce{^{22}Ne}, assuming the free value of
$g_s^{p}$ ($g_s^{n}$) and an effective charge of $e_{p} = 1.5e$ ($e_{n} =0.5e$) for the protons (neutrons). In contrast, for the $2_1^+$ state
of \ce{^{26}Mg} and \ce{^{30}Si}, less agreement with the central values of the experimental data is attained \cite{SaS17}. In both cases, 
a more satisfactory answer is provided by the IM-SRG approach, that yields $g=+0.512$ and $g=+0.420$ for the two nuclides, respectively \cite{SaS17}, 
both more compatible with the experimental values.

Regarding \ce{^{34}S}, the g-factor obtained in the large-scale SM calculation with \textit{sd}-\textit{pf} configuration mixing in ref.~\cite{WiC79}
is in excellent agreement with the observed result, whose central value is the closest to the macroscopic $\alpha$-cluster value 
and its statistical error is smaller than for the neighbouring nuclei in tab.~\ref{T-2-05}. The mixing between the \textit{sd} and \textit{pf}
major shells has not been considered in the \ce{^{38}Ar} case, where the g-factor of the $2_1^+$ is obtained entirely from the \textit{sd} model 
space with the Warburton-Brown interactions \cite{WaB92}, but with the same convention for the nucleon charges as well as for $g_s^{n}$ and $g_s^{p}$. 
The result is still compatible with the experimental g-factor (cf. tab.~\ref{T-2-05}).

The large-scale \textit{sd}-\textit{pf} SM calculation in ref.~\cite{SHS03} produces a compatible value for the g-factor of 
the $2_1^+$ state of \ce{^{42}Ca}, which the Schmidt estimate in tab.~\ref{T-2-05} reproduces with the same magnitude but opposite sign.
Underpinning this calculation is the enlarged model space, that assumes the $\alpha$-conjugate nucleus \ce{^{28}Si} as a core and the USD 
interactions \cite{Wil84} in the \textit{sd}-shell, the modified Kuo-Brown ones \cite{PoZ81} in the \text{pf}-shell and the LKS potentials \cite{KLS69} 
between the two major shells. The \ce{^{40}Ca} core excitations are found to provide substantial improvement with respect to the \textit{pf}-shell
estimates reported in ref.~\cite{SHS03}, thus indicating a deviation from sphericity in the \ce{^{40}Ca} core. 
Eventually, the \textit{pf}-shell
model (FSM) calculation in ref.~\cite{ESK00} with modified Kuo-Brown (KB3) interactions \cite{PoZ81} sensibly underestimates the g-factor 
of the lowest $2^+$ state of \ce{^{46}Ti}, as a result of the limited size of the model space. In \ce{^{45}Ti} (\ce{^{46}Ti}), the smallness of the excitation energy 
of the $7/2_1^-$ ($2_1^+$) state suggests an underlying \textit{collective} nature. Hence, a larger number of valence nucleons in the SM 
calculation or an $\alpha$-clustered \ce{^{44}Ti} core would be recommendable.


\section{\textsf{The \ce{^{12}C} nucleus}}\label{S-3.0}

Here, we present the theoretical predictions for the magnetic dipole moment of two energy levels of the
\ce{^{12}C} nucleus within a macroscopic $\alpha$-cluster model. For the latter, we select the Hamiltonian
with the $\alpha$-$\alpha$ interaction
given by the isotropic Ali-Bodmer \cite{AlB66} potential in ref.~\cite{FKK04} with the same parameters as
in sec.~II~A of ref.~\cite{LLL14}. On top 
of the latter, reproducing the short-range repulsive and long-range attractive effects of the strong
force (cf. eqs.~(2) and (3) of ref.~\cite{SEM18}),
we add the Coulomb interaction, accounting for the spherical charge distribution of the \ce{^4He} nucleus
with charge radius $R_{\alpha} = 1.44$~fm
(cf. eq.~(4) of ref.~\cite{SEM18}), as well as the Gaussian $3\alpha$ force (cf. eq.~(5) of
ref.~\cite{SEM18}), whose strength and range parameters
were originally fitted to reproduce the binding energy of \ce{^{12}C} and the $2_1^+$ - $0_2^+$
energy difference respectively (cf. refs.~\cite{PoC79,FKK04}).
However, due to the adopted isotropic Ali-Bodmer potential with the parameters given in ref.~\cite{LLL14},
the energy of the ground state of this nucleus 
coincides with the opposite of the Hoyle state gap in the continuum and infinite-volume limit. 

As in refs.~\cite{LLL14,SEM18}, we transpose the above system of $\alpha$-particles in a finite
cubic lattice of side $L = Na$ with $N$ points
per dimension, with $a$ the lattice spacing. To the wavefunctions of the lattice Hamiltonian
(cf. sec.~3.1 and 3.2 of ref.~\cite{SEM18}), we impose 
periodic boundary conditions (PBCs). As a consequence, the $M$-body relative configuration space in
the continuum and infinite volume, 
$\mathbb{R}^{3M-3}$, is reduced to a torus in $3M$-$3$ dimensions. This operation produces manifold
implications in the Hamiltonian and its eigenfunctions,
 the most glaring of them is rotational symmetry breaking, represented by the descent in the
 symmetry from SO(3) 
to the cubic group, $\mathrm{SO}(3,\mathbb{Z})$ \cite{LLL14,LLL15} or $\mathcal{O}$ \cite{SEM18,Ste20},
consisting of the 24 rotations of the regular hexahedron. 

The latter and the other discrete symmetries of the lattice Hamiltonian, outlined in sec.~4
of ref.~\cite{SEM18}, permit to classify the lattice eigenstates 
in terms of the irreducible representations of parity, $\mathcal{O}$ and the group of the permutation of
$M$ particles. Taking
the bosonic nature of $\alpha$-particles into account, we consider only the completely symmetric
representation of the latter group,
\begin{equation}
{\tiny\yng(2)\ldots \tiny\yng(1)} = [\mathrm{M}] \equiv S ~,\label{E-3.0-01}
\end{equation}
and apply the associated projectors in the numerical procedure for the extraction of the
lattice eigenstates (cf. sec.~6 of ref.~\cite{SEM18}).
As in refs.~\cite{LLL14,SEM18}, we introduce the lattice counterpart of SO(2), i.e. the cyclic
group of order four, $\mathcal{C}_4$, associated 
to the counterclockwise rotation of $\pi/2$ about the $z$ axis. Labeling the irreps of the former,
$I_z$, by positive integers modulo 4 \cite{LLL14}, 
and identifying the irreps of the cubic group with $\Gamma$ \cite{SEM18}, we represent each lattice
eigenstate with a round bracket
 (cf. ref.~\cite{LLL15}) as 
\begin{equation}
|\mathscr{N}, \Gamma, I_z , S, \mathscr{P} ),\label{E-3.0-02}
\end{equation}
where $\mathscr{N}$ is a positive nonzero integer that denotes the order with which the lattice
eigenstates with the same transformation 
properties appear in the spectrum, thus fulfilling the role of the principal quantum number.
When the SO(3) multiplet to which each state belongs 
in the continuum and infinite volume is identified, the angular momentum quantum number, $\ell$,
is added to the lattice states. Starting 
from states with well-defined $I_z$ (cf. eq.~\eqref{E-3.0-02}) and part of the same SO(3) multiplet,
it is possible to obtain states with
good angular momentum projection along the z-axis in the continuum and infinite volume limit.
The replacement of the $\mathcal{C}_4$ by the 
SO(2) label is performed through the unitary transformation discussed in app.~C of
ref.~\cite{Ste20} and presented for completeness
in the tabs.~\ref{T-A-03}-\ref{T-A-14} for $\ell \leq 9$ in the appendix. In this transformed basis,
the lattice eigenstates become
\begin{equation}
|\mathscr{N}, \ell , m , S, \mathscr{P} ),\label{E-3.0-03}
\end{equation}
where the label $\Gamma$ has been dropped for brevity. 

Exactly the states in eq.~\eqref{E-3.0-03} constitute the main ingredient for the computation of the
average values of the magnetic dipole 
moment operator (cf. \cite{RiS04}), defined in the relative reference frame as
\begin{equation}
\begin{split}
\mu(\mathscr{N}, \ell, \mathscr{P})^{(r)} \equiv 
( \mathscr{N}~\ell~\ell~S~\mathscr{P} | \hat{\mu}_0^{(r)} | \mathscr{N}~\ell~\ell~S~\mathscr{P} ) \\
= ( \mathscr{N}~{\ell}~{\ell}~S~\mathscr{P} |  \mu_N g_{L}^{(\alpha)} \sum_{i = 1}^{M-1} 
\frac{(\mathcal{L}_{iM})_z}{\hbar} & |  \mathscr{N}~{\ell}~{\ell}~S~\mathscr{P} )\label{E-3.0-04}
\end{split}
\end{equation}
where $g_{L}^{(\alpha)}$ is given by eq.~\eqref{E-2.1-02} and $(\mathcal{L}_{iM})_z$ is the $z$-component
of the angular momentum operator on the lattice.
Its expression, consistently with the formula of the squared total angular momentum derived in
sec.~3.3 of ref.~\cite{SEM18}, is defined in the relative frame by
\begin{equation}
(\mathcal{L}_{iM})_z = a \hbar\sum_{\mathbf{n}_i\in \mathscr{N}}
\sum_{k=1}^{K} C_k^{(1,K)} (\mathbf{n}_{iM})_{x} 
\left[a_{iM}^{\dagger}(\mathbf{n}_{iM} +  k\mathbf{e}_{y})a_{iM}(\mathbf{n}_{iM}) 
- a_{iM}^{\dagger}(\mathbf{n}_{iM}-k\mathbf{e}_{y})a_{iM}(\mathbf{n}_{iM})\right]~,\label{E-3.0-05}
\end{equation}
where $a_{iM}^{\dagger}(\mathbf{n}_{iM})$ and $a_{iM}(\mathbf{n}_{iM})$ are the creation and
annihilation operators of the $\alpha$-particle $i$ relative to
the $\alpha$-particle $M$ at the site $(\mathbf{n}_{iM})$ of the lattice, whereas the coefficients
$C_k^{(1,K)}$, that scale as the inverse of the lattice spacing, 
are defined in refs.~\cite{SEM18,Ste20}.
Considering the fact that the center-of-mass degrees of freedom have been dropped from
the Ha\-miltonian, the l.h.s. of eq.~\eqref{E-3.0-04}
is equivalent to the magnetic dipole moment in the absolute reference frame,
\begin{equation}
\begin{split}
\mu(\mathscr{N}, \ell, \mathscr{P})^{(a)} \equiv 
( \mathscr{N}~\ell~\ell~S~\mathscr{P} | \hat{\mu}_0^{(a)} | \mathscr{N}~\ell~\ell~S~\mathscr{P} )\\ 
= ( \mathscr{N}~{\ell}~{\ell}~S~\mathscr{P} |  \mu_N g_{L}^{(\alpha)} 
\sum_{i = 1}^{M} \frac{(\mathcal{L}_i)_z}{\hbar} & |  \mathscr{N}~{\ell}~{\ell} 
~S~\mathscr{P} )~,\label{E-3.0-06}
\end{split}
\end{equation}
 where the coordinate of the $M^{th}$ particle is extracted from the relative ones of the other
 particles, subject to the constraint that  $\mathbf{r}_{CM} = 0$.

Furthermore, the computation of the magnetic dipole moment associated with the two energy levels,
provides the opportunity to continue the analysis of the artifacts induced by the lattice environment
in average values of spherical tensor operators that started in ref.~\cite{LLL15}.
In this respect, we here focus on the behaviour of the average values of the former observable with
the lattice spacing and, aiming at reducing the 
discretization errors, we compute the \textit{isotropic averages} (ref.~\cite{LLL15}) of the
magnetic dipole moments.
Extending the definition in eq.~(21) of ref.~\cite{LLL15} to the magnetic dipole moment in the
relative frame in eq.~\eqref{E-3.0-04},
the isotropically averaged expression of $\mu(\mathscr{N}, \ell, \mathscr{P})^{(r)}$ is obtained,
\begin{equation}
\begin{split}
( \mathscr{N}~\ell~\ell~S~\mathscr{P} | \hat{\mu}_0^{(r)} | \mathscr{N}~\ell~\ell~S~\mathscr{P} )_{\circ} 
= (\ell 1 \ell|\ell 0\ell ) \frac{1}{2\ell+1} \sum_{m,m',m''}(\ell 1 \ell| m m' m'') \\ \cdot ( \mathscr{N}~\ell~m''~S~
\mathscr{P}|  \mu_N g_{L}^{(\alpha)}\sum_{i=1}^{M-1} \frac{(\mathcal{L}_{iM})_{m'}}{\hbar} & |\mathscr{N}~\ell~m~S~\mathscr{P})~,\label{E-3.0-07}
\end{split}
\end{equation}
where the lattice angular momentum operator is expressed in the spherical basis and the Clebsch-Gordan
coefficents of SO(3), $(\ell \ell' \ell''|m m' m'')$, have 
been introduced in the notation of ref.~\cite{MaG96}. 

The counterpart of the last equation in the absolute reference frame can be obtained
by applying the same definition (cf. ref.~\cite{LLL15}) to the matrix element 
$( \mathscr{N}~\ell~\ell~S~\mathscr{P} | \hat{\mu}_0^{(a)} | \mathscr{N}~\ell~\ell~S~\mathscr{P} )_{\circ}$.
Additionally, in the three sums over the SO(2) labels at the r.h.s. of eq.~\eqref{E-3.0-07} only the
projections $m$ and $m''$ differing by zero or one 
unit of $\hbar$ provide a nonzero contribution. To ponder the amount of each of the latter to
the related isotropic average, it is convenient to
define approximate reduced matrix elements, by resorting to the Wigner-Eckart theorem for SO(3)
(cf. eq.~(2.169) in ref.~\cite{MaG96}). 
Applying the theorem, we define the reduced matrix elements of $\hat{\mu}$ between lattice eigenstates,
\begin{equation}
 ( \ell~m || \hat{\mu}_{m'}^{(\rho)} || \ell~m'' ) \equiv 
 \frac{( \mathscr{N}~\ell~m~S~\mathscr{P} |  \hat{\mu}_{m'}^{(\rho)} 
| \mathscr{N}~\ell~m''~S~\mathscr{P} )}{(\ell 1 \ell| m'' m' m)} ~, \label{E-3.0-08}
\end{equation}
in the relative and in the absolute reference frames, $\rho\equiv a$ or $r$. The definition in
eq.~\eqref{E-3.0-08} coincides with the one  given in eqs.~(15) and (19) of ref.~\cite{LLL15}, up
to a factor equal to the Clebsch-Gordan coefficient $(\ell 1 \ell| 0 0 0)$ multiplied by the
constant $\sqrt{3/4\pi}$. 

As in the latter study, due to the breaking of rotational symmetry, the $\mathrm{SO}(2)$ labels
have not been dropped in the definition of the reduced 
brackets in eq.~\eqref{E-3.0-07}.  Additionally, we assume that the Clebsch-Gordan coefficients on
the r.h.s. of eq.~\eqref{E-3.0-08} do not 
vanish, a fact that is guaranteed by the triangular inequality between the angular momenta,
$|\ell - 1| \leq \ell \leq \ell + 1$ and the conservation 
of the third component of the angular momentum $m' + m'' = m$. Besides, the latter three projections
are also required to be nonzero, since the sum of the three angular momenta in eq.~\eqref{E-3.0-08},
$2\ell + 1$, is odd \cite{MaG96}. 

Reduced matrix elements can be immediately extended to the isotropic averages (cf. eq.~\eqref{E-3.0-07}),
\begin{equation}
 ( \ell~m || \hat{\mu}_{m'}^{(\rho)}  || \ell~m'' )_{\circ}  \equiv 
 \frac{( \mathscr{N}~\ell~m~S~\mathscr{P} | \hat{\mu}_{m'}^{(\rho)} | \mathscr{N}~\ell~m''~S~\mathscr{P} )_{\circ}}{(\ell 1 \ell| m'' m' m)}~,\label{E-3.0-09}  
\end{equation}
with the same convention on the $\rho$ index. Both the quantities in eqs.~\eqref{E-3.0-08} and \eqref{E-3.0-09} are expected to reproduce 
asymptotically the exact reduced matrix elements in the continuum and infinite-volume limit,
\begin{equation}
 \langle \ell || \hat{\mu}^{(\rho)} || \ell \rangle  \equiv 
 \frac{\langle N~\ell~m~S~\mathscr{P} | \hat{\mu}_{m'}^{(\rho)} | N~\ell~m''~S~\mathscr{P} \rangle}{(\ell 1 \ell| m'' m' m)},~\label{E-3.0-10}    
\end{equation}
where $N$ is the counterpart of $\mathscr{N}$ in the $\mathbb{R}^{3M-3}$ configuration space.

\begin{figure}[ht!]
\includegraphics[width=1.0\columnwidth]{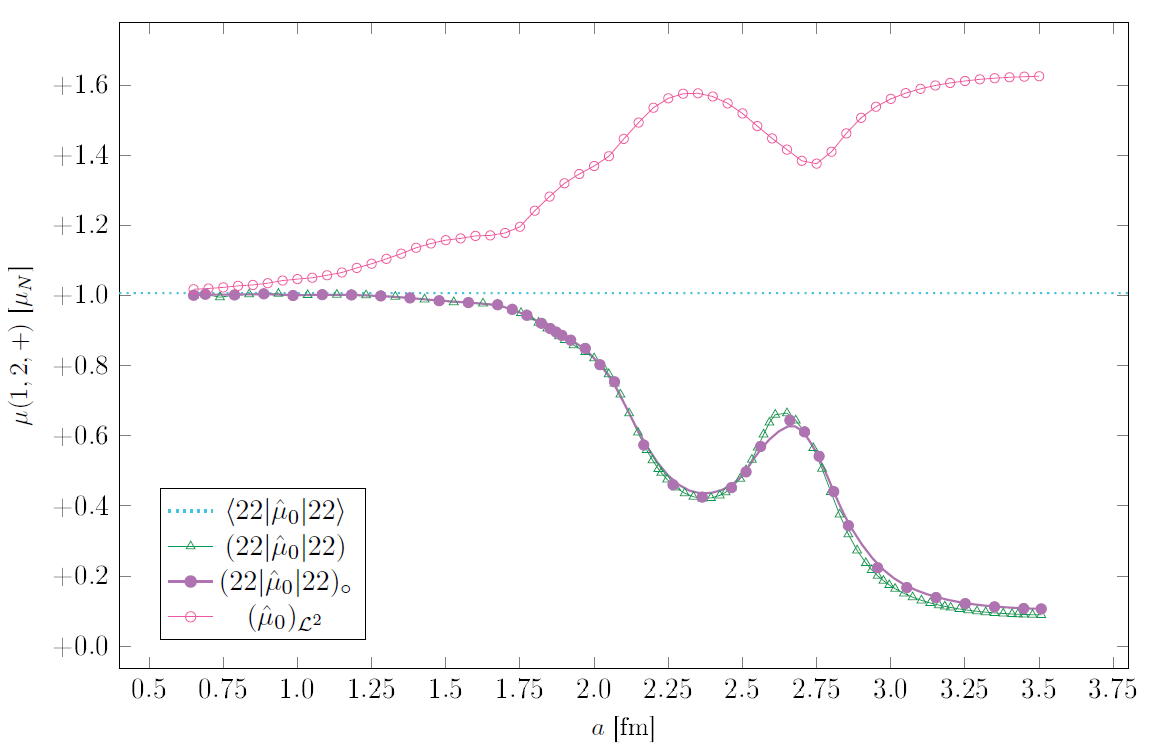}
\caption{Average value of the magnetic dipole moment for the $2_1^+$ energy level of \ce{^{12}C} as
a function of the lattice spacing. The broken line with 
circles denotes the estimated magnetic dipole moments obtained from the multiplet-averaged values of
the squared total angular momentum operator, 
$\mathcal{L}^2$ \cite{SEM18,Ste20}, over the five wavefunctions composing the $2_1^+$ level,
degenerate in the continuum and infinite volume limit. 
The broken line with triangles denotes the matrix elements of the magnetic dipole moment operator
(cf. eqs.~\eqref{E-3.0-04} or \eqref{E-3.0-06}) 
over the lattice states with maximum projection along the $z$ axis. The isotropically-averaged estimates of the same observable (cf. eq.~\eqref{E-3.0-07}),
extracted from the matrix element of  $\hat{\boldsymbol{\mu}}$ with lattice wavefunctions with
a different third component of the total angular 
momentum operator, are displayed by the broken lines with squares. The theoretical value
of $\approx +1.0068~\mu_N$ is marked with a dotted line and reproduced
with $10^{-3}$ precision by the values of $\langle 2 2 |\hat{\mu}_0 | 2 2 \rangle$ and
$( 2 2 |\hat{\mu}_0 | 2 2 )_{\circ}$ at $a \approx 0.65$~fm, both equal to $\approx 1.0018~\mu_N$.
Finite-volume effects are suppressed by the constraint $Na \geq 19$~fm. }
\label{F-3-01}
\end{figure}

Let us now begin with the behaviour of the magnetic dipole moment of the $2_1^+$ state at
$4.4398(2)$~MeV of \ce{^{12}C} as a function of the lattice spacing, 
displayed in fig.~\ref{F-3-01}. As a result of the tuning of the parameters of the
phenomenological $2\alpha$ and $3\alpha$ potentials of the present model \cite{LLL14,SEM18}, 
in the continuum and infinite-volume limit the energy eigenvalue of this state converges 
to $\approx 3.35$~MeV, as shown by the curves associated with the $E$ and $T_2$
multiplets of the cubic group and by the corresponding 
multiplet-averaged (cf. eq.~(50) of ref.~\cite{SEM18}) solid line in figs.~28 and~32 in
ref.~\cite{SEM18}. By setting the side of the cubic lattice to 
$L \geq 19$~fm as in refs.~\cite{LLL14,SEM18} we have reduced the finite-volume errors
associated with the energy eigenvalues to $\approx$ $10^{-2}$~MeV, 
in order to remove the latter artifacts in the present analysis of discretization effects.

In its generality, the expression of the magnetic dipole moment operator for $\alpha$ conjugate
nuclei in eq.~\eqref{E-2.1-01} permits to obtain average values 
for $\hat{\mu}$ among states with maximum angular momentum projection, starting from estimates of
the maximum $\hat{L}_z$ eigenvalue. In the present case, the latter 
can be extracted from the average values of the squared total angular momentum operator as a function
of the lattice spacing analyzed in fig.~39 of ref.~\cite{SEM18}. 
Exploiting eq.~\eqref{E-2.1-01} and the expression of the eigenvalues of $\mathcal{L}^2$
(cf. eqs.~(24)-(27) of ref.~\cite{SEM18}) in the continuum and infinite volume, 
$\ell(\ell + 1)\hbar^2$, we have obtained the mulitplet-averaged estimates of the magnetic dipole
moments for the $2_1^+$ state, that we denote with $(\hat{\mu}_0)_{\mathcal{L}^2}$. 
In fig.~\ref{F-3-01}, the open circles at different values of the lattice spacing indeed
describe a curve that lies always above the expectation value 
of the magnetic dipole moment at $+1.0068~\mu_N$, marked by a dotted line in fig.~\ref{F-3-01}.

In particular, the minimum in $(\hat{\mu}_0)_{\mathcal{L}^2}$
at $a \approx 2.80$~fm corresponds to a deep minimum of the average value of $\mathcal{L}^2$ in the
$E$ multiplet of $\mathcal{O}$,
constituting the $2_1^+$ energy state in the continuum and infinite-volume limit (cf. fig.~(39) of
ref.~\cite{SEM18}). In contrast to energy eigenvalues and the 
average values of the $\alpha$-$\alpha$ separation (cf. $\mathcal{E}_r$ and $\mathcal{R}$ in
figs.~32 and~33 of ref.~\cite{SEM18}), the minima of the total
squared angular momentum operator and hence $(\hat{\mu}_0)_{\mathcal{L}^2}$ cannot be unambigously
mapped to the local maxima of the probability density function associated to the lattice
eigenstates they refer to.
Concerning the isotropic average of the magnetic dipole moment (cf. eq.~\eqref{E-3.0-07}), its behaviour
with the lattice spacing shown by the curve
with solid circles in fig.~\ref{F-3-01}, follows quite faithfully the one of the magnetic dipole moment
(open triangles in fig.~\ref{F-3-01}) computed from 
the lattice $2_1^+$ eigenstates with maximum angular momentum projection. 

More precisely, $(22|\hat{\mu}_0|22)_{\circ}$ slightly quenches the discretization artifacts in the very
large lattice spacing region ($a \gtrsim 2.75$~fm)
 as well as in a small region around $a\approx 2.37$~fm. 
However, for $a \lesssim 2.20$~fm no significant improvement in the estimate of the magnetic moment
from the isotropic average is observed, 
whereas in the peak region around ($a \approx 2.60$~fm) the deviation from the asymptotic value of
$\mu(1,2,+)$ becomes even larger than for 
$(22|\hat{\mu}_0|22)$.

\begin{figure}[ht!]
\includegraphics[width=1.0\columnwidth]{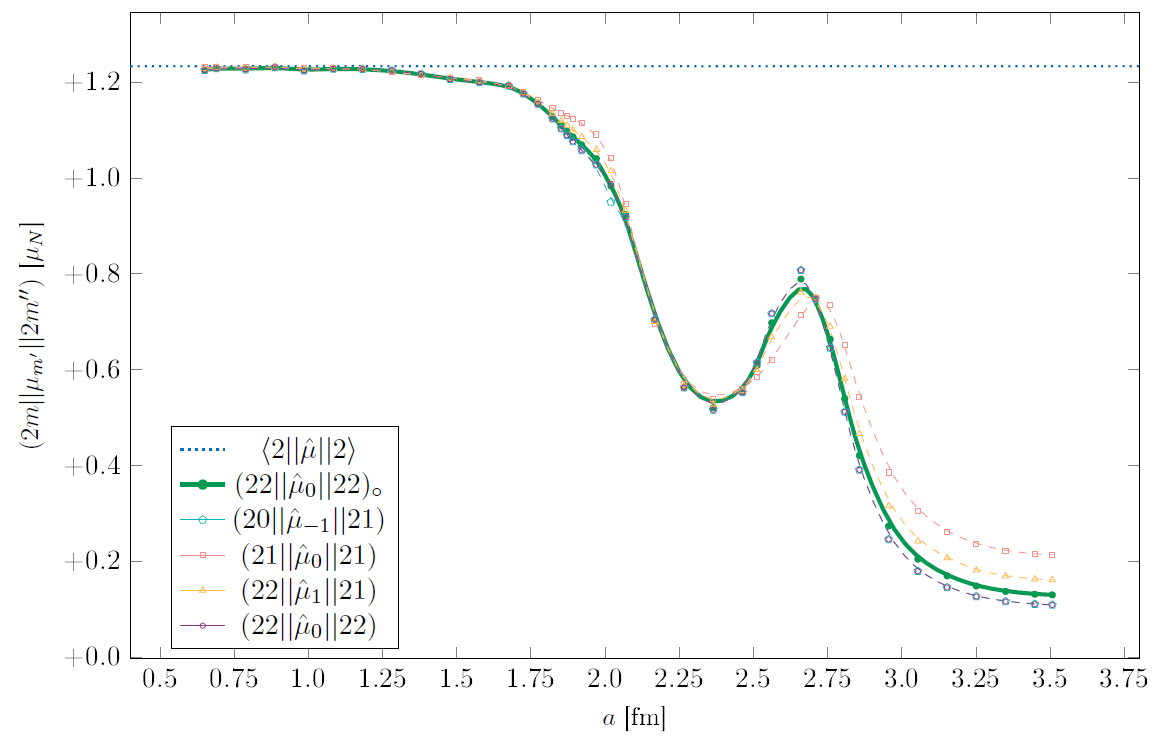}
\caption{Behaviour of the reduced matrix elements between the spherical components of the magnetic
dipole moment operator and lattice eigenstates with 
different angular momentum projection, $m$ and $m''$, along the $z$-axis as a function of the
lattice spacing $a$ for the $2_1^+$ state of \ce{^{12}C}. 
The listed matrix elements represent the nonvanishing independent contributions to the
isotropically-averaged estimate of the magnetic dipole moment in eq.~(18), 
divided by the Clebsch-Gordan coefficient $(212|mm'm'')$. The asymptotic value of the reduced
matrix elements in the continuum and infinite volume 
limit is represented by a dotted line and is independent on the third component of the angular
momentum. The curves underlying each sample of matrix 
elements denote basis-spline functions calculated on the points of each dataset. In particular,
the dashed (solid and thick) curve with open (filled) circles 
represents the reduced (isotropically-averaged) magnetic dipole moment. The diagonal
matrix elements $( 2 1 \lvert \lvert \hat{\mu}_{0} \lvert \lvert 2 1 )$ 
with intermediate projection are denoted by a dashed curve with open squares, lying above all
other curves in the large lattice spacing region. Finally, the 
behaviour of the off-diagonal matrix elements  $( 2 0 \lvert \lvert \hat{\mu}_{-1} \lvert \lvert 2 1 )$
and $( 2 2 \lvert \lvert \hat{\mu}_{1} \lvert \lvert 2 1 )$ 
is described by the curves with open pentagons and triangles, respectively.}
\label{F-3-02}
\end{figure}

In summary, the isotropic average slightly reduces the discrepancies in the
magnetic dipole moment between the extremal points.
The reasons underlying this behaviour can be better understood by considering the individual
reduced matrix elements (cf. eq.~\eqref{E-3.0-08}) 
contributing to the reduced isotropic average in eq.~\eqref{E-3.0-09}. As it can be proven by
rewriting the angular momentum operator in spherical components, the 
matrix elements contributing to the isotropic average in eq.~\eqref{E-3.0-07} are overall
12 and fulfill the following symmetry relations:
\begin{equation}
( 2 0 \lvert \lvert \hat{\mu}_{0} \lvert \lvert 2 0 ) = 0~,\nonumber
\end{equation}
that vanishes as the associated Clebsch-Gordan coefficient, and 
\begin{equation}
( 2 2 \lvert \lvert \hat{\mu}_{0} \lvert \lvert 2 2 ) = - ( 2 -2 \lvert \lvert \hat{\mu}_{0} \lvert \lvert 2 -2 )~,\nonumber
\end{equation}
\begin{equation}
( 2 1 \lvert \lvert \hat{\mu}_{0} \lvert \lvert 2 1 ) = - ( 2 -1 \lvert \lvert \hat{\mu}_{0} \lvert \lvert 2 -1 )~,\nonumber
\end{equation}
\begin{equation}
( 2 0 \lvert \lvert \hat{\mu}_{-1} \lvert \lvert 2 1 ) = -( 2 1 \lvert \lvert \hat{\mu}_{1} \lvert \lvert 2 0 ) 
= ( 2 -1 \lvert \lvert \hat{\mu}_{-1} \lvert \lvert 2 0 )  = -( 2 0 \lvert \lvert \hat{\mu}_{1}  \lvert \lvert 2 -1)~,\nonumber
\end{equation}
and
\begin{equation}
( 2 -1 \lvert \lvert \hat{\mu}_{1} \lvert \lvert 2 -2 ) = -( 2 -2 \lvert \lvert \hat{\mu}_{-1} \lvert \lvert 2 -1 ) \\
= ( 2 2 \lvert \lvert \hat{\mu}_{1} \lvert \lvert 2 1 ) = -( 2 1 \lvert \lvert \hat{\mu}_{-1}  \lvert \lvert 2 2)~.\label{E-3.0-11}  
\end{equation}
Thanks to the identities in eq.~\eqref{E-3.0-11}, in fig.~\ref{F-3-02} we show the contributions of
four reduced matrix elements 
to $( 2 2 \lvert \lvert \hat{\mu}_{1} \lvert \lvert 2 2 )_{\circ}$ in total. In particular, the elements
$( 2 0 ||\hat{\mu}_{-1} || 2 1 )$ 
and $( 2 2 ||\hat{\mu}_{0} || 2 2 )$ display almost overlapping paths (cf. dashed lines with
open pentagons and circles, respectively, in fig.~\ref{F-3-02}), with 
slight deviations in the neighbourhood of $a\sim2.05$~fm. Besides, these two contributions
lie closest to the isotropic average and share the extrema with the 
latter, denoted by the solid curve with circles (cf. fig.~\ref{F-3-02}). 

Conversely, the dashed curve with open squares in fig.~\ref{F-3-02}, that identifies the
diagonal matrix element with intermediate angular momentum projection, 
displays overall the largest discrepancies from the isotropic average and exhibts a displaced
local maximum at $a \approx 2.75$~fm, a fact does not uniquely depend on different the 
Clebsch-Gordan coefficients that multiply the various contributions to the isotropic average.

Similarly, the curve for the reduced matrix element $( 2 1 || r^{2} Y_{2}^{0} || 2 1 )$, 
associated with the electric qua\-drupole moment of the $2_1^+$ state of \ce{^{8}Be} in fig.~2 of
ref.~\cite{LLL15} \footnote{The matrix elements 
of the spherical tensor operators discussed in ref.~\cite{LLL15} are expressed in unit of charge.}
displays the largest deviation in the shape and in 
the magnitude with respect to the relevant isotropic average (cf. the solid line with filled triangles
in the latter figure). The same observation can be 
drawn from the matrix element $( 2 1 || r^{4} Y_{4}^{0} || 2 1 )$ for the electric hexadecupole moment
of the same state of \ce{^{8}Be} in fig.~3 
of ref.~\cite{LLL15}. In the latter graph, also the contribution from the diagonal element with
maximum angular momentum projection highlights large 
deviation from the isotropic average in magnitude, but the position of the extrema is closer
to the one of the isotropic average than in the  $( 2 1 || r^{4} Y_{4}^{0} || 2 1 )$ case. 

Moreover, from fig.~\ref{F-3-02} it can be inferred that the curve of non-diagonal matrix element
$( 2 2 ||\hat{\mu}_{1} || 2 1 )$ follows a path 
intermediate between the one of the isotropic average and the $( 2 1 ||\hat{\mu}_{0} || 2 1 )$
element, with a slightly shifted position of the local maximum.

Finally, from comparison between the magnetic dipole moment in the $2_1^+$ of \ce{^{12}C} in
fig.~\ref{F-3-02} and the quadrupole and hexadecupole 
electric moments in the $2_1^+$ state of \ce{^{8}Be} in figs.~2 and 3 of ref.~\cite{LLL15},
we can conclude that the former observable is significantly 
less affected by discretization errors, in the same range of lattice spacing. More precisely,
while all reduced matrix elements 
$(2 m'' || \hat{\mu}_{m'} || 2 m)$ at $a\approx 1.75$~fm differ by less than 10\% to the continuum
and infinite volume counterpart, at the same lattice 
spacing some of the contributions to the isotropically-averaged electric qudrupole and
hexadecupole moment differ by more than 50 \%
from their respective asymptotic values.

Second, we consider the magnetic dipole moment of the $3_1^-$ state at 9.641(5)~MeV of \ce{^{12}C}.
In the adopted macroscopic $\alpha$-cluster approach, 
the energy eigenvalue of the lattice counterpart of the $3_1^-$ state coverges to $\approx$~5.85~MeV
above the \textit{g.s.} in the continuum and infinite-volume limit (cf. fig.~36 of ref.~\cite{SEM18}).

\begin{figure}[ht!]
\includegraphics[width=1.0\columnwidth]{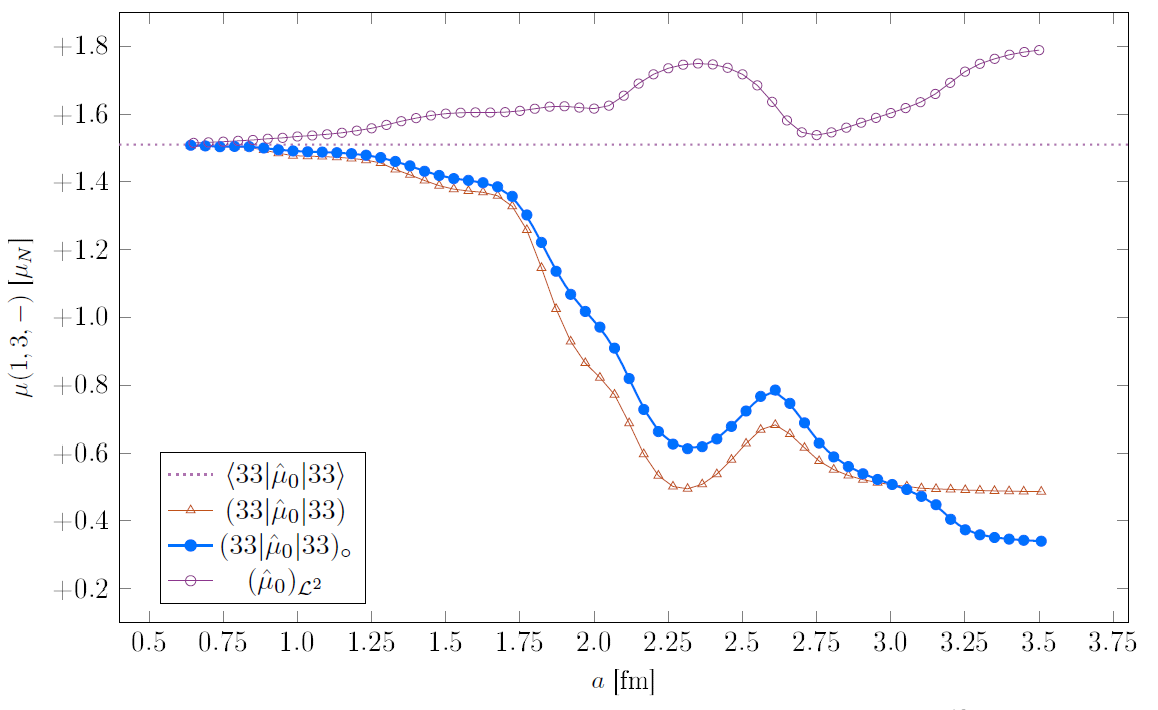}
\caption{Average value of the magnetic dipole moment for the $3_1^-$ energy level of \ce{^{12}C}
as a function of the lattice spacing. The broken line with 
circles denotes the estimated magnetic dipole moments obtained from the multiplet-averaged values
of the squared total angular momentum operator, 
$\mathcal{L}^2$ \cite{SEM18,Ste20}, over the five wavefunctions composing the $3_1^-$ level,
degenerate in the continuum and infinite volume limit. 
The broken line with triangles denotes the matrix elements of the magnetic dipole moment operator 
over the lattice states with maximum projection along the z axis. The isotropically-averaged estimates
of the same observable, extracted from the matrix element of  $\hat{\boldsymbol{\mu}}$ with lattice
wavefunctions with different third component of the total angular 
momentum operator, are displayed by the broken lines with squares. The theoretical value at
$1.5102~\mu_M$ is marked with a dotted line and reproduced
with $10^{-3}$ precision by the values of $\langle 3 3 |\hat{\mu}_0 | 3 3 \rangle$ and
$( 3 3 |\hat{\mu}_0 | 3 3 )_{\circ}$ at $a \approx 0.64$~fm, equal to $\approx 1.5057$ and 
$1.5081~\mu_N$ respectively. Finite-volume effects are suppressed by the constraint $Na \geq 19.5$~fm.}
\label{F-3-03}
\end{figure}

Starting from the multiplet-averaged value of the squared total angular momentum for the lowest
$A_2^-$, $T_1^-$ and $T_2^-$ states (cf. fig.~40 of ref.~\cite{SEM18}), 
an estimate of magnetic dipole moment can be obtained as for the $2_1^+$ state. Fixing the
lattice size $L \gtrsim 19.5$~fm, the corresponding magnetic moments, $(\hat{\mu}_0)_{\mathcal{L}^2}$,
have been evaluated in an interval of lattice spacing ranging from $0.65$ to $3.50$~fm. As shown
by the curve with open circles in fig.~\ref{F-3-03}, these estimates lie above the expectiation
value of the magnetic moment (dotted line), with a minimum at $a\approx 2.80$~fm. The extremum is
found in good correspondence with a sharp minimum in the average value of the squared
angular momentum of the $T_1$ multiplet of states (cf. fig.~40 of ref.~\cite{SEM18}). 

Concerning the magnetic dipole moment computed from the state with maximum angular momentum
projection along the $z$ axis, $(3 3| \mu_0| 33)$, the curve with open triangles in fig.~\ref{F-3-03}
reaches the theoretical value of $\mu(1,3,-)$ in the continuum limit from below and displays a
minimum at $\approx 2.29$~fm, as the curve for the $2_1^+$ level of \ce{^{12}C} (cf. fig.~\ref{F-3-01}).
However, the decrease of $(3 3| \mu_0| 33)$ in the region $1.75 \lesssim a \lesssim 2.25$~fm is
steeper than the one of $(2 2| \mu_0| 22)$ in the same interval (cf. fig.~\ref{F-3-01}). Additionally,
the discretization artifacts on $(3 3| \mu_0| 33)$ at $a \gtrsim 1.75$~fm are significantly
larger than for $(\hat{\mu}_0)_{\mathcal{L}^2}$, in contrast with fig.~\ref{F-3-01}. 

\begin{figure}[ht!]
\includegraphics[width=1.0\columnwidth]{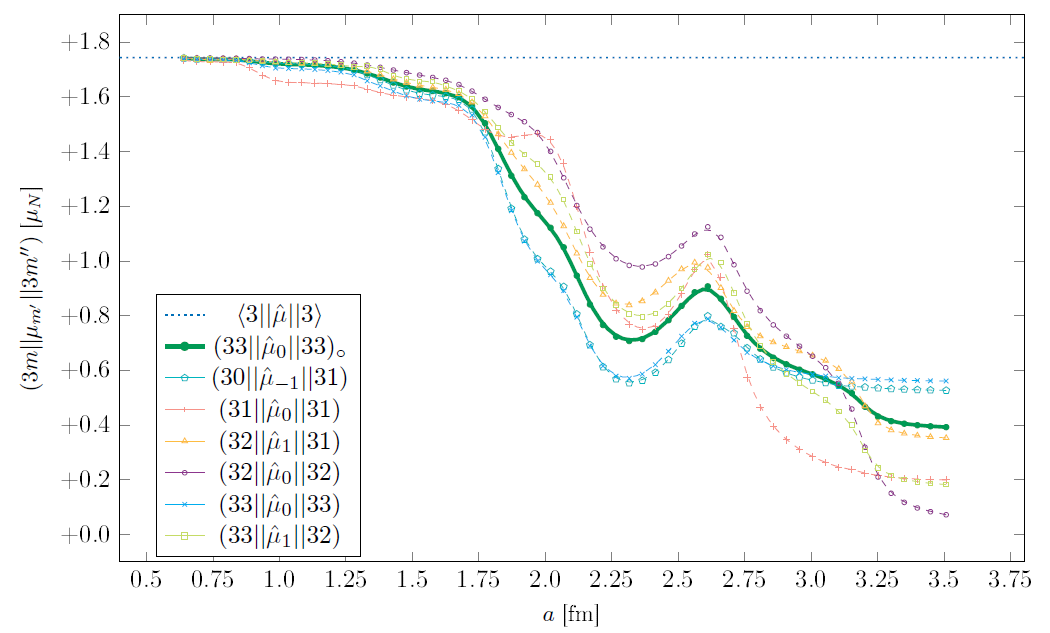}
\caption{Behaviour of the reduced matrix elements between the spherical components of the magnetic
dipole moment operator and lattice eigenstates with 
different angular momentum projection, $m$ and $m''$, along the $z$-axis as a function of the
lattice spacing $a$ for the $3_1^-$ state of \ce{^{12}C}. 
The listed matrix elements represent the nonvanishing independent contributions to the
isotropically-averaged estimate of the magnetic dipole moment in eq.~(18), 
divided by the Clebsch-Gordan coefficient $(313|mm'm'')$. The asymptotic value of the reduced
matrix elements in the continuum and infinite volume 
limit is represented by a dotted line and is independent on the third component of the
angular momentum. In particular, the dashed (solid and thick) curve with open (filled) circles 
represents the reduced (isotropically-averaged) magnetic dipole moment. The diagonal matrix elements
$( 3 2 \lvert \lvert \hat{\mu}_{0} \lvert \lvert 3 2 )$ 
and $( 3 1 \lvert \lvert \hat{\mu}_{0} \lvert \lvert 3 1 )$ 
with intermediate projection are denoted by a dashed curve with vertical and diagonal crosses,
lying above all other curves in the large lattice spacing region. Finally, the 
behaviour of the off-diagonal matrix elements  $( 3 0 \lvert \lvert \hat{\mu}_{-1} \lvert \lvert 3 1 )$,
$( 3 2 \lvert \lvert \hat{\mu}_{1} \lvert \lvert 3 1 )$ and 
$( 3 3 \lvert \lvert \hat{\mu}_{1} \lvert \lvert 3 2 )$ is described by lines with open pentagons,
triangles and squares, respectively.}
\label{F-3-04}
\end{figure}

Nonetheless, the calculation of the isotropicaly-averaged value of $(3 3| \mu_0| 33)$
(cf. eq.~\eqref{E-3.0-07}) yields a sizable suppression of the discretization errors
in the region $1.80 \lesssim a \lesssim 2.80$~fm, with a peak of $\sim 15\%$ around the local
minimum at $a\approx 2.35$~fm. The curve for $(3 3| \mu_0| 33)_{\circ}$, indeed, lies above the one
for $(3 3| \mu_0| 33)$ in the whole interval $ 0.65 \lesssim a \lesssim 3.10$~fm
(cf. fig.~\ref{F-3-03}), unlike the $2_1^+$ case (cf. fig.~\ref{F-3-01}).
The origin of this behaviour can be better investigated by considering the individual
contributions (cf. eq.~\eqref{E-3.0-08}) to the reduced isotropic average in eq.~\eqref{E-3.0-09}.
Specifically, the non-vanishing matrix elements on the r.h.s. of eq.~\eqref{E-3.0-07} are
now 18 and fulfill the following symmetry identities:
\begin{equation}
( 3 0 \lvert \lvert \hat{\mu}_{0} \lvert \lvert 3 0 ) = 0~,\nonumber
\end{equation}
\begin{equation}
( 3 3 \lvert \lvert \hat{\mu}_{0} \lvert \lvert 3 3 ) = - ( 3 -3 \lvert \lvert \hat{\mu}_{0} \lvert \lvert 3 -3 )~,\nonumber
\end{equation}
\begin{equation}
( 3 2 \lvert \lvert \hat{\mu}_{0} \lvert \lvert 3 2 ) = - ( 3 -2 \lvert \lvert \hat{\mu}_{0} \lvert \lvert 3 -2 )~,\nonumber
\end{equation}
\begin{equation}
( 3 1 \lvert \lvert \hat{\mu}_{0} \lvert \lvert 3 1 ) = - ( 3 -1 \lvert \lvert \hat{\mu}_{0} \lvert \lvert 3 -1 )~,\nonumber
\end{equation}
\begin{equation}
( 3 0 \lvert \lvert \hat{\mu}_{-1} \lvert \lvert 3 1 ) = -( 3 1 \lvert \lvert \hat{\mu}_{1} \lvert \lvert 3 0 ) 
= ( 3 -1 \lvert \lvert \hat{\mu}_{-1} \lvert \lvert 3 0 )  = -( 3 0 \lvert \lvert \hat{\mu}_{1}  \lvert \lvert 3 -1)~,\nonumber
\end{equation}
\begin{equation}
( 3 -1 \lvert \lvert \hat{\mu}_{1} \lvert \lvert 3 -2 ) = -( 3 -2 \lvert \lvert \hat{\mu}_{-1} \lvert \lvert 3 -1 ) 
= ( 3 2 \lvert \lvert \hat{\mu}_{1} \lvert \lvert 3 1 ) = -( 3 1  \lvert \lvert \hat{\mu}_{-1}  \lvert \lvert 3 2)~,\nonumber 
\end{equation}
and
\begin{equation}
( 3 -2 \lvert \lvert \hat{\mu}_{1} \lvert \lvert 3 -3) = -( 3 -3 \lvert \lvert \hat{\mu}_{-1} \lvert \lvert 3 -2 ) 
= ( 3 3 \lvert \lvert \hat{\mu}_{1} \lvert \lvert 3 2 ) = -( 3 2 \lvert \lvert \hat{\mu}_{-1}  \lvert \lvert 3 3)~.\label{E-3.0-11}  
\end{equation}
Due to the latter relations, the independent contributions to $(33||\hat{\mu}_1||33)_{\circ}$ reduce to
six in total (cf. fig.~\ref{F-3-04}). 
Differently from the $2_1^+$ multiplet in fig.~\ref{F-3-02}, the behaviour of these reduced
brackets as a function of the lattice spacing is more multifaceted. 
In particular, the diagonal matrix element with maximum angular momentum projection (cf. the curve
with diagonal crosses in fig.~\ref{F-3-04})
follows a path that almost overlaps the one of the element $(3 0||\hat{\mu}_{-1}||31)$ (pentagons)
throughout the considered interval of lattice spacing,
displaying a local minimum at $a\approx 2.33$~fm, followed by a local maximum at $a\approx2.67$~fm
and a plateau in the large lattice-spacing region.

In the opposite direction, the two matrix elements undergo a steep increase with an
inflection point at $a\approx1.90$~fm, that eventualy results in a plateau 
in the continuum limit. Both $(3 0||\hat{\mu}_{-1}||31)$ and $(3 3||\hat{\mu}_{0}||33)$ are  most affected  by discretization errors 
in the region $1.90 \lesssim a \lesssim 2.75$~fm. 

Conversely, the diagonal matrix element with projection
$2$ (open circles) shows the smallest deviation 
from the exact matrix element, $\langle 3 || \hat{\mu} || 3 \rangle$ (dotted line), in the whole domain,
except for a region $3.20 \lesssim a \lesssim 3.27$~fm. Regarding the $2_1^+$ state, this fact remains
true only in the interval $2.48 \lesssim a \lesssim 2.75$~fm (cf. open circles in fig.~\ref{F-3-02}).
However, in the two cases, the diagonal matrix elements with projection $2$ drop to $\approx 0.10
\mu_N$ at $a \approx 3.5$~fm. 

A behaviour intermediate between the one of the $(3 3||\hat{\mu}_{0}||33)$ and $(3 2||\hat{\mu}_{0}||32)$
curves is found in the off-diagonal matrix elements $(33||\hat{\mu}_1||32)$ (open squa\-res) and
$(32||\hat{\mu}_1||31)$ (open triangles), which indeed lie closest to the isotropic average
(solid curve with full circles),
except for intersection regions at $1.85$ and $3.18$~fm and in the stretch between the two local
extrema. An inflection point is also detected around $a\approx 3.20$~fm, as for the
$(3 2||\hat{\mu}_{0}||32)$ curve (cf. fig.~\ref{F-3-04}).

Concerning the diagonal matrix element with projection $1$ (vertical crosses), its curve displays
a second local maximum at $a\approx 1.97$~fm, followed by another minimum at $a\approx 1.88$~fm,
in contrast with all other reduced matrix elements. Additionally, its convergence to the continuum
and infinite-volume
value of $(3 1||\hat{\mu}_{0}||31)$ is the slowest among the brackets of the magnetic dipole moment
operator, since $1 \%$ precision is reached only at $a \approx 0.85$~fm. 

However, the two minima of the curve of the $(3 1||\hat{\mu}_{0}||31)$ bracket admit an
interpretation on the basis of the range parameters $\eta_1^{-1}$ and $\eta_{0}^{-1}$ of the
Ali-Bodmer potential \cite{FKK04}, equal to $1.89$ and $2.29$~fm respectively. The probability
density function associated with the lattice state $| 3 1)$, in fact, possesses local maxima
that can be included in the lattice points when $a \approx \eta_1^{-1}$ (cf. sec.~2 of ref.~\cite{SEM18}).

\section{\textsf{Conclusion}}\label{S-2.4}

With the present work, we have complemented the vast literature on magnetic dipole moments
\cite{CaT90,ZHG78,BRA77}, by paralleling the average values of this 
observable for even-even self-conjugate nuclei from \ce{^{12}C} to \ce{^{44}Ti} with the
partitioning of nuclear matter into $\alpha$-clusters.
The measured gyromagnetic factors of excited states of these nuclei turn out to assume,
within their errors, the same value of $g\approx +0.50$.
It is shown that this specific feature can be explained on the basis of collective
excitations of \ce{^{4}He} nuclei, neglecting the degrees of 
freedom of the single nucleons. Such macroscopic approaches are best suited to low-lying
excited states located at the vicinity of $\alpha$-decay thre\-sholds. 

The stability of the $\alpha$-clusters has been further investigated for isotopes of
the above chain of nuclei by appending one or two neutrons.
Relying on Schmidt estimates \cite{Sch37}, we confirmed that, in proximity of the shell
closures $Z,N = 8$ and $20$ the extra one or two neutrons determine the 
experimental g-factor values. The original positive values of the corresponding self-conjugate
nuclei undergo a significant reduction generated by the contribution of the negative g-factor
of the additional neutrons. For the considered energy levels of the $N = Z +1$ and $2$ semi-magic
nuclei in our chains, $\alpha$-clustering 
is mitigated by the added neutrons. Outside the shell closures, for $N = Z + 2$ nuclei the
added neutrons loose almost all their influence,
and the predictions of eq.~\eqref{E-2.1-02} are aligned with the experimental g-factors.

Spurred by the recent literature on the subject \cite{LLL14,LLL15,SEM18,SLL22}, we have calculated
the magnetic dipole moment of the $2_1^+$ state at
$4.44$~MeV and the $3_1^-$ state at $9.64$~MeV of \ce{^{12}C} in the framework of a
macroscopic $\alpha$-cluster model on the lattice in ref.~\cite{LLL14}. 
Suppressing the finite-volume effects, we have analyzed the discretization artifacts through
the behaviour of this observable as a function of the lattice spacing,
ranging from $a \approx 0.65$ to $3.50$~fm.

Unlike the energy (cf. fig.~32 in ref.~\cite{SEM18}) or the squared angular momentum
(cf. fig.~39 in ref.~\cite{SEM18} and the 
curve with open circles in fig.~\ref{F-3-01}), the magnetic dipole moment appears rather
insensitive to discretization effects for $a \lesssim 2.0$~fm,
where the artifacts amount to, at most, the $15\%$ of the continuum and inifinite-volume value,
governed by eq.~\eqref{E-2.1-01}. 

In particular, for the $2_1^+$ state, the isotropic average, introduced in ref.~\cite{LLL15},
turns out not to improve significantly the estimate of the 
magnetic dipole moment provided by matrix element with maximum angular momentum projection along
the lattice $z$-axis alone (cf. the curve 
with solid circles in fig.~\ref{F-3-01}). The latter conclusion does not hold in the $3_1^-$ case,
where the isotropic average turns out to improve sizably the esitmate of this observable in
the region with $a \lesssim 3.05$~fm (cf. fig.~\ref{F-3-03}). 

If its weak coupling to angular momentum symmetry breaking is confirmed in further nuclear
systems on the lattice, the magnetic dipole 
moment evaluated at finite and sizable values of the lattice spacing (\textit{e.g.} $a \sim 2.0$~fm)
can corroborate the classification of lattice eigenstates
in terms of $\mathrm{SO}(3)$ quantum numbers, which is exact only in the continuum and
infinite-volume limit. 


\section*{\textsf{Acknowledgements}}
We thank T. Otsuka for the shell-model calculation of the g-factors of the $2_1^+$ states of 
$^{10}\mathrm{Be}$ and $^{10 }\mathrm{C}$.
We gratefully acknowledge funding from the Deutsche Forschungsgemeinschaft
(DFG, German Research Fo\-undation) and the NSFC through the funds provided
to the Sino-German Collaborative Research Center TRR110 ``Symmetries
and the Emergence of Structure in QCD'' (DFG Project ID 196253076
- TRR 110, NSFC Grant No. 12070131001), the Chinese Academy of Sciences
(CAS) President's International Fellowship Initiative (PIFI) (Grant
No. 2018DM0034), Volkswagen Stiftung (grant No. 93562),  the Euro\-pean
Research Council (ERC) under the European Union's Horizon 2020 research
and innovation programme (grant agreement No. 101018170) and the Espace de Structure Nucléaire Théorique (ESNT) of the CEA/DSM-DAM. The computational 
resources of the "CLAIX" cluster of the RWTH in Aachen ('jara0015' project, No. 24986) as well as
the ones of the "Jean Zay" cluster of the IDRIS institute (CNRS, GENCI) in Orsay 
(DARI 'dqt' project,  No. 102105/A0110513012) have been exploited.


\begin{appendices}

\section*{\textsf{Appendix}}\label{S-A}

We here recapitulate the basic properties of the cubic group and the transformation tables for basis states
 of irreducible representations (irreps) of $\mathrm{SO}(3)$ with $\ell \leq 9$ into the ones of $\mathcal{O}$ (cf. ref.~\cite{Ste20}).

\begin{table}[ht!]
\begin{center}
\scalebox{0.90}{ 
{\renewcommand\arraystretch{1.2} 
\begin{tabular}{ccccc}
\toprule
 E & $6C_2''$ & $3C_4^2(\pi)$ & $8C_3'$ & $6C_4(\frac{\pi}{2})$ \\
\midrule
$(0,0,0)$ & $(0,\pi,\frac{\pi}{2})$ & $(\pi,\pi,0)$ & $(\frac{\pi}{2},\frac{\pi}{2},\pi)$ & $(\frac{\pi}{2},\frac{\pi}{2},\frac{3\pi}{2})$\\
& $(0,\pi,\frac{\pi}{2})$ & $(0,\pi,0)$ & $(\pi,\frac{3\pi}{2},\frac{3\pi}{2})$ & $(\frac{3\pi}{2},\frac{\pi}{2},\frac{\pi}{2})$\\
& $(0,\pi,\frac{3\pi}{2})$ & $(\pi,0,0)$ & $(\pi,\frac{3\pi}{2},\frac{\pi}{2})$ & $(\pi,\frac{\pi}{2},\pi)$\\
& $(\frac{3\pi}{2},\frac{\pi}{2},\frac{3\pi}{2})$ & & $(\frac{3\pi}{2},\frac{\pi}{2},\pi)$ & $(\pi,\frac{3\pi}{2},\pi)$\\
& $(0,\frac{\pi}{2},\pi)$ & & $(\pi,\frac{\pi}{2},\frac{3\pi}{2})$ & $(\frac{\pi}{2},0,0)$\\
& $(\pi,\frac{\pi}{2},0)$ & & $(\frac{\pi}{2},\frac{3\pi}{2},\pi)$ & $(\frac{3\pi}{2},0,0)$\\
& & & $(\pi,\frac{\pi}{2},\frac{\pi}{2})$ & \\
& & & $(\frac{3\pi}{2},\frac{3\pi}{2},\pi)$ & \\
\bottomrule
\end{tabular}}}
\end{center}
\caption{Representation of the group. The elements of each conjugacy
 class are listed in terms of Euler angles. For completeness, the symmetry operation $(\alpha, \beta,\gamma)$
 consists of a rotation of angle $\gamma$ about the $z$-lattice axis, followed by one of
 angle $\beta$ about the $y$ axis and by another of angle $\alpha$ about the $z$ axis.}\label{T-A-01}
\end{table}

This group consists of 24 rotations about the symmetry axes of the regular hexahedron and the octahedron, 
partitioned into five equivalence classes. Adopting the notation by Sch\"onflies \cite{Car97}, $E$ represents the
 identity, $3C_4^2(\pi)$ the rotations of $180^{\circ}$ about the three fourfold axes orthogonal
 to the faces of the cube, $6C_4(\pi/2)$ the $90^{\circ}$ and $270^{\circ}$
 rotations about the latter axes (6 elements), $6C_4''$ the $180^{\circ}$ rotations about the six diagonal
 axes parallel to two faces of the cube and $8C_3'(2\pi/3)$ are rotations of $120^{\circ}$ and $240^{\circ}$
 about the four diagonal axes that cross two opposite vertexes of the cube (8 elements).\\
Furthermore, the characters of the five irreducible representations of $\mathcal{O}$ are presented
 in tab.~\ref{T-A-01}, together with the characters of $2\ell+1$-dimensional irreps of $\mathrm{SO}(3)$.  

Applying the great orthogonality theorem, the decomposition of the $2\ell+1$-dimensional representations of the 
rotation group into the $\#\mathcal{C}l$ cubic group irreps, is obtained (cf. tab.~\ref{T-A-02} and tab.~II 
of ref.~\cite{LLL14}),
\begin{equation}
D^{\ell} = \sum_{\oplus} q_{\nu} D^{\nu}~,\label{E-A-01}
\end{equation}
where the multiplicity of the latter into the original irrep $D^{\ell}$ is denoted by
\begin{equation}
q_{\nu} = \frac{1}{|\mathcal{O}|} \sum_{i = 1}^{\#\mathcal{C}l}
|\mathcal{C}l_i|[\chi_i^{\nu}]^*\chi_i^{\ell}\label{E-A-02}
\end{equation}
In the last equation, the order of $\mathcal{O}$ of the finite group is at the denominator, whereas $\chi_i^{\nu}$ and $\chi_i^{\ell}$ 
represent respectively the characters of the irreps of the cubic and the rotation group in the conjugacy class $\mathcal{C}l_i$ with 
$|\mathcal{C}l_i|$ elements. 
Moreover, the map between the basis states of $\mathrm{SO}(3,\mathbb{Z})$ and $\mathrm{SO}(3)$ irreps 
is provided by the projectors in eq.~(40) of ref.~\cite{SEM18}. \\

\begin{table}[h!]
\begin{center}
\scalebox{0.90}{
{\renewcommand\arraystretch{1}
\begin{tabular}{c|ccccc}
\toprule
$\Gamma$ & E & $6C_2''$ & $3C_4^2(\pi)$ & $8C_3'$ & $6C_4(\frac{\pi}{2})$ \\
\midrule
$A_1$ & 1 & 1 & 1 & 1 & 1\\
$A_2$ & 1 & -1 & 1 & 1 & -1 \\
$E$ & 2 & 0 & 2 & -1 & 0 \\
$T_1$ & 3 & -1 & -1 & 0 & 1 \\
$T_2$ & 3 & 1 & -1 & 0 & -1 \\
$D^{\ell}$ & $2\ell+1$ & $(-1)^{\ell}$ & $(-1)^{\ell}$ & $1-\mathrm{mod}(\ell,3)$ & $(-1)^{[\frac{\ell}{2}]}$ \\
\bottomrule
\end{tabular}}}
\end{center}
\caption{Character table of the cubic group. With the exception of the $\ell=0,1$ cases,
 this representation $D^{\ell}$ is fully reducible with respect to the $\mathcal{O}$ operations.}\label{T-A-02}
\end{table}

Let  $T_q^{(k)}$ denote the $q$ component of a spherical 
tensor of rank $2k+1$, the general component of the irreducible cubic tensor obtained from it is 
\begin{equation}
T_q^{(\Gamma,k)} = \sum_{q' = -k}^k
 \sum_{g \in\mathcal{O}}\chi_{\Gamma}(g)D_{qq'}^{k}(g) T_{q'}^{(k)}\label{E-A-03}
\end{equation}
where the label $k$ in the spherical tensor on the l.h.s. denotes the original irrep of $\mathrm{SO}(3)$ from which it has been 
obtained, whereas the index $q$ ranges from $-k$ to $k$. In contrast, the transpose transformation rule holds 
for the basis states of the two groups, 
\begin{equation}
|\ell, \Gamma, m\rangle = \sum_{m' = -\ell}^{\ell} \sum_{g \in\mathcal{O}}
\chi_{\Gamma}(g)D_{m'm}^{\ell}(g) |\ell,m'\rangle.\label{E-A-04}
\end{equation}
As a consequence of this descent in symmetry, in the cubic lattice environment the maximum rank of any 
irreducible tensor operator runs from one to three. As shown in sec.~4 of ref.~\cite{SEM18} for the energy eigenstates,
 the non-null components $q$ of  $T^{(\Gamma,k)}$ and $|\Gamma \ell \rangle$, admixture
 of the $q \mod 4$ components of their $\mathrm{SO}(3)$ counterparts, can be unambigously denoted with the
 $I_z$ label. The ensuing distribution of $m$ components of a spin-$\ell$ irrep into
 the $(\ell, \Gamma)$ irreps of $\mathrm{SO}(3, \mathbb{Z})$ is known as \emph{subduction} \cite{DEP09}.

Furthermore, when the multiplicity coefficient $q_{\Gamma}$ of the irrep $\Gamma$ of 
$\mathcal{O}$ is larger than one, further linear transformations on the outcoming states 
(cf. eq.~\eqref{E-A-03}) or cubic tensor components (cf. eq.~\eqref{E-A-04}) 
should be performed, in order to block-diagonalize the relevant projector
 and disentangle the repeated multiplets of states \cite{Ste20}. 

Eventually, for the construction of the tables (cf. tab.~\ref{T-A-04}-\ref{T-A-14}), the basis 
eigenfunctions for the '0' and '2' irreps of  the cyclic group of order four, $\mathcal{C}_4$, generated by an element of the conjugacy class 
$6C_4(\pi/2)$ of the cubic group (e.g. the counterclockwise rotation of $\pi/2$ about the $z$-axis of the cube) are 
assumed to be real, whereas the following phase convention
\begin{equation}
\left(
\begin{array}{c}
\Psi_{\Gamma,I_z = 1}\\
\Psi_{\Gamma,I_z = 3}\\
\end{array}
\right)
= -\frac{1}{\sqrt{2}}
\left(
\begin{array}{c}
\Psi_{\Gamma}^{(p)} + \mathrm{i}  \Psi_{\Gamma}^{(q)}\\
- \Psi_{\Gamma}^{(p)} + \mathrm{i}   \Psi_{\Gamma}^{(q)}
\end{array}
\right).\label{E-A-05}
\end{equation}
 for the basis state $\Psi_{\Gamma,I_z = 1}$ ($\Psi_{\Gamma,I_z = 3}$) belonging to the '1' ('3') irrep of $\mathcal{C}_4$ is understood.\\

\begin{table}[h!]
\begin{minipage}[c]{.21\columnwidth}
\centering
\scalebox{0.95}{
\begin{tabular}{cc|c}
\toprule
 & $\ell$ & $ 0$\\
$\Gamma$ & \backslashbox{$I_z$}{$m$} & 0\\
\midrule
$A_1$ & 0 & $1$\\
\bottomrule
\end{tabular}}
\end{minipage}
\begin{minipage}[c]{.33\columnwidth}
\centering
\scalebox{0.90}{
\begin{tabular}{cc|ccc}
\toprule
 & $\ell$ & \multicolumn{3}{c}{$1$}\\
$\Gamma$ & \backslashbox{$I_z$}{$m$} & $-1$ & $0$ & $1$\\
\midrule
\multirow{3}{0.25cm}{\centering{$T_1$}} & $0$ & & $1$ &\\
& $1$ & &  & $1$ \\
& $3$ & $1$ & & \\
\bottomrule
\end{tabular}}
\end{minipage}
\begin{minipage}[c]{.46\columnwidth}
\centering
\scalebox{0.90}{
\begin{tabular}{cc|ccccc}
\toprule
 & $\ell$ & \multicolumn{5}{c}{$2$}\\
$\Gamma$ & \backslashbox{$I_z$}{$m$} & $-2$ & $-1$ & $0$ & $1$ & $2$\\
\midrule
\multirow{2}{0.25cm}{\centering{$E$}} & $0$ & & & $1$ & &\\
& $2$ & $\sqrt{1/2}$ & & & & $\sqrt{1/2}$\\
\midrule
\multirow{3}{0.25cm}{\centering{$T_2$}} & $1$ & & & &  $1$ &\\
& $2$ & $\mathrm{i}\sqrt{1/2}$ & & & & $-\mathrm{i}\sqrt{1/2}$\\
& $3$ &  & $1$ & & &\\
\bottomrule
\end{tabular}}
\end{minipage}
\caption{Decomposition tables for basis states of $\mathrm{SO}(3)$ with $\ell \leq 2$ into irreps of $\mathcal{O}$.}\label{T-C-04}
\end{table}

\bigskip

\begin{table}[h!]
\begin{center}
\scalebox{0.90}{
\begin{tabular}{cc|ccccccc}
\toprule
 & $\ell$ & \multicolumn{7}{c}{$3$}\\
$\Gamma$ & \backslashbox{$I_z$}{$m$} & $-3$ & $-2$ & $-1$ & $0$ & $1$ & $2$ & $3$\\
\midrule
$A_2$ & $2$ & & $\mathrm{i}\sqrt{\frac{1}{2}}$ & & & & -$\mathrm{i}\sqrt{\frac{1}{2}}$ & \\
\midrule
\multirow{3}{0.25cm}{\centering{$T_1$}} & $0$ & & & & $1$ & & &\\
& $1$ & $\sqrt{\frac{5}{8}}$ & & & & $\sqrt{\frac{3}{8}}$ & &\\
& $3$ & & & $\sqrt{\frac{3}{8}}$ & & & & $\sqrt{\frac{5}{8}}$\\
\midrule
\multirow{3}{0.25cm}{\centering{$T_2$}} & $1$ & $\sqrt{\frac{3}{8}}$ & & & & $-\sqrt{\frac{5}{8}}$ & &\\
& $2$ & & $\sqrt{\frac{1}{2}}$ & & & & $\sqrt{\frac{1}{2}}$ & \\
& $3$ & & & $-\sqrt{\frac{5}{8}}$ & & & & $\sqrt{\frac{3}{8}}$\\
\bottomrule
\end{tabular}
}
\end{center}
\caption{Decomposition table for basis states of $\mathrm{SO}(3)$ with $\ell = 3$ into irreps of $\mathcal{O}$.}\label{T-C-05}
\end{table}

\bigskip
\begin{center}
\scalebox{0.99}{
\begin{tabular}{cc|ccccccccc}
\toprule
 & $\ell$ & \multicolumn{9}{c}{$4$}\\
$\Gamma$ & \backslashbox{$I_z$}{$m$} & $-4$ & $-3$ & $-2$ & $-1$ & $0$ & $1$ & $2$ & $3$ & $4$\\
\midrule
$A_1$ & $0$ & $\frac{1}{2}\sqrt{\frac{5}{6}}$ & & & & $\frac{1}{2}\sqrt{\frac{7}{3}}$ & & & & $\frac{1}{2}\sqrt{\frac{5}{6}}$\\
\midrule
\multirow{2}{0.25cm}{\centering{$E$}} & $0$ & $\frac{1}{2}\sqrt{\frac{7}{6}}$ & & & & -$\frac{1}{2}\sqrt{\frac{5}{3}}$ & & & & $\frac{1}{2}\sqrt{\frac{7}{6}}$\\
& $2$ & & & $\sqrt{\frac{1}{2}}$ & & & & $\sqrt{\frac{1}{2}}$ & &\\
\midrule
\multirow{3}{0.25cm}{\centering{$T_1$}} & $0$ & $\mathrm{i}\sqrt{\frac{1}{2}}$ & & & & & & & & -$\mathrm{i}\sqrt{\frac{1}{2}}$\\
& $1$ &  & $\frac{1}{2}\sqrt{\frac{1}{2}}$ & & & & $\frac{1}{2}\sqrt{\frac{7}{2}}$ & & &\\
& $3$ & & & & $\frac{1}{2}\sqrt{\frac{7}{2}}$ & & & & $\frac{1}{2}\sqrt{\frac{1}{2}}$ &\\
\midrule
\multirow{3}{0.25cm}{\centering{$T_2$}} & $1$ &  & $\frac{1}{2}\sqrt{\frac{7}{2}}$ & & & & -$\frac{1}{2}\sqrt{\frac{1}{2}}$ & & &\\
& $2$ &  & & $\mathrm{i}\sqrt{\frac{1}{2}}$ & & & & -$\mathrm{i}\sqrt{\frac{1}{2}}$ & &\\
& $3$ & & & & -$\frac{1}{2}\sqrt{\frac{1}{2}}$ & & & & $\frac{1}{2}\sqrt{\frac{7}{2}}$ &\\
\bottomrule
\end{tabular}}
\captionof{table}{Decomposition table for basis states of $\mathrm{SO}(3)$ with $\ell = 4$ into irreps of $\mathcal{O}$.}\label{T-A-05}
\end{center}

\scalebox{0.85}{
\begin{tabular}{cc|ccccccccccc}
\toprule
 & $\ell$ & \multicolumn{11}{c}{$5$}\\
$\Gamma$ & \backslashbox{$I_z$}{$m$} & $-5$ & $-4$ & $-3$ & $-2$ & $-1$ & $0$ & $1$ & $2$ & $3$ & $4$ & $5$\\
\midrule
\multirow{2}{0.25cm}{\centering{$E$}} & $0$ & & $\mathrm{i}\sqrt{\frac{1}{2}}$ & & & & & & & & -$\mathrm{i}\sqrt{\frac{1}{2}}$ &\\
& $2$ & & & & $\mathrm{i}\sqrt{\frac{1}{2}}$ & & & & -$\mathrm{i}\sqrt{\frac{1}{2}}$ & & &\\
\midrule
\multirow{6}{0.25cm}{\centering{$T_1$}} & $0$ & & $\frac{1}{2}\sqrt{\frac{7}{6}}$ & & & & $\frac{1}{2}\sqrt{\frac{5}{3}}$ & & & & $\frac{1}{2}\sqrt{\frac{7}{6}}$ & \\
& $1$ &  & & -$\frac{1}{4}\sqrt{\frac{7}{6}}$ & & & & $\frac{3}{4}$ & & & & $\frac{1}{4}\sqrt{\frac{35}{6}}$\\
& $3$ & $\frac{1}{4}\sqrt{\frac{35}{6}}$ & & & & $\frac{3}{4}$ & & & & -$\frac{1}{4}\sqrt{\frac{7}{6}}$ & &\\
& $0$ & & $\frac{1}{2}\sqrt{\frac{5}{6}}$ & & & & -$\frac{1}{2}\sqrt{\frac{7}{3}}$ & & & & $\frac{1}{2}\sqrt{\frac{5}{6}}$ & \\
& $1$ &  & & $\sqrt{\frac{5}{6}}$ & & & & & & & & $\sqrt{\frac{1}{6}}$\\
& $3$ & $\sqrt{\frac{1}{6}}$ & & & & & & & & $\sqrt{\frac{5}{6}}$ & &\\
\midrule
\multirow{3}{0.25cm}{\centering{$T_2$}} & $1$ &  & & $\frac{1}{4}\sqrt{\frac{3}{2}}$ & & & & $\frac{\sqrt{7}}{4}$ & & & & -$\frac{1}{4}\sqrt{\frac{15}{2}}$\\
& $2$ &  & & & $\sqrt{\frac{1}{2}}$ & & & & $\sqrt{\frac{1}{2}}$ & & &\\
& $3$  & -$\frac{1}{4}\sqrt{\frac{15}{2}}$ & & & & $\frac{\sqrt{7}}{4}$ & & & & $\frac{1}{4}\sqrt{\frac{3}{2}}$ & &\\
\bottomrule
\end{tabular}}
\captionof{table}{Decomposition table for basis states of $\mathrm{SO}(3)$ with $\ell = 5$ into irreps of $\mathcal{O}$.}\label{T-A-06}
\smallskip
\begin{center}
\scalebox{0.97}{
\begin{tabular}{cc|ccccccc}
\toprule
 & $\ell$ & \multicolumn{7}{c}{$6$}\\
$\Gamma$ & \backslashbox{$I_z$}{$m$} & $-6$ & $-5$ & $-4$ & $-3$ & $-2$ & $-1$ & $0$\\
\midrule
$A_1$ & $0$ & & & $\frac{\sqrt{7}}{4}$ & & & & $-\sqrt{\frac{1}{8}}$ \\
\midrule
$A_2$ & $2$ & $\frac{1}{4}\sqrt{\frac{5}{2}}$ & & & & $-\frac{1}{4}\sqrt{\frac{11}{2}}$ & &\\
\midrule
\multirow{2}{0.25cm}{\centering{$E$}} & $0$ & & & $\frac{1}{4}$ & & & & $\frac{1}{2}\sqrt{\frac{7}{2}}$ \\
& $2$ & $\frac{1}{4}\sqrt{\frac{11}{2}}$ & & & & $\frac{1}{4}\sqrt{\frac{5}{2}}$ & &\\
\midrule
\multirow{3}{0.25cm}{\centering{$T_1$}} & $0$ & & & $\mathrm{i}\sqrt{\frac{1}{2}}$ & & & & \\
& $1$ & & & & $\frac{1}{4}\sqrt{\frac{15}{2}}$ & & & \\
& $3$ & & $\frac{1}{4}\sqrt{\frac{11}{2}}$ & & & & $-\frac{\sqrt{3}}{4}$ &\\
\midrule
\multirow{6}{0.25cm}{\centering{$T_2$}} & $1$ & & & & $-\frac{1}{4}\sqrt{\frac{55}{14}}$ & & & \\
& $2$ & $\mathrm{i}\frac{1}{4\sqrt{7}}$ & & & & $\mathrm{i}\frac{1}{4}\sqrt{\frac{55}{7}}$ & &\\
& $3$  &  & $\frac{1}{4}\sqrt{\frac{21}{2}}$ & & & & $\frac{1}{4}\sqrt{\frac{11}{7}}$ &\\
& $1$ & & & & $\sqrt{\frac{2}{7}}$ & & & \\
& $2$ & $\mathrm{i}\frac{1}{4}\sqrt{\frac{55}{7}}$ & & & & -$\mathrm{i}\frac{1}{4\sqrt{7}}$ & &\\
& $3$  &  & & & & & $\sqrt{\frac{5}{7}}$ &\\
\bottomrule
\end{tabular}
}
\captionof{table}{Decomposition table for basis states of $\mathrm{SO}(3)$ with $\ell = 6$ and $m \leq 0$ into irreps of $\mathcal{O}$.}\label{T-A-07}
\end{center}

\begin{center}
\scalebox{0.85}{
\begin{tabular}{cc|cccccc}
\toprule
 & $\ell$ & \multicolumn{6}{c}{$6$}\\
$\Gamma$ & \backslashbox{$I_z$}{$m$} & $1$ & $2$ & $3$ & $4$ & $5$ & $6$\\
\midrule
$A_1$ & $0$  & & & & $\frac{\sqrt{7}}{4}$ & &\\
\midrule
$A_2$ & $2$  & & $-\frac{1}{4}\sqrt{\frac{11}{2}}$ & & & & $\frac{1}{4}\sqrt{\frac{5}{2}}$\\
\midrule
\multirow{2}{0.25cm}{\centering{$E$}} & $0$  & & & & $\frac{1}{4}$ & &\\
& $2$ & & $\frac{1}{4}\sqrt{\frac{5}{2}}$ & & & & $\frac{1}{4}\sqrt{\frac{11}{2}}$\\
\midrule
\multirow{3}{0.25cm}{\centering{$T_1$}} & $0$  & & & & $-\mathrm{i}\sqrt{\frac{1}{2}}$ & & \\
& $1$ &  $-\frac{\sqrt{3}}{4}$ & & & & $\frac{1}{4}\sqrt{\frac{11}{2}}$ & \\
& $3$ & & & $\frac{1}{4}\sqrt{\frac{15}{2}}$ & & &\\
\midrule
\multirow{6}{0.25cm}{\centering{$T_2$}} & $1$  & $\frac{1}{4}\sqrt{\frac{11}{7}}$ & & & & $\frac{1}{4}\sqrt{\frac{21}{2}}$ & \\
& $2$  & & $-\mathrm{i}\frac{1}{4}\sqrt{\frac{55}{7}}$ & & & & $-\mathrm{i}\frac{1}{4\sqrt{7}}$\\
& $3$  & & & $-\frac{1}{4}\sqrt{\frac{55}{14}}$ & & &\\
& $1$  & $\sqrt{\frac{5}{7}}$ & & & & & \\
& $2$  & & $\mathrm{i}\frac{1}{4\sqrt{7}}$ & & & & -$\mathrm{i}\frac{1}{4}\sqrt{\frac{55}{7}}$\\
& $3$  & & & $\sqrt{\frac{2}{7}}$ & & &\\
\bottomrule
\end{tabular}}
\captionof{table}{Decomposition table for basis states of $\mathrm{SO}(3)$ with $\ell = 6$ and $m >0$ into irreps of $\mathcal{O}$.}\label{T-A-08}
\end{center}
\smallskip
\begin{center}
\scalebox{0.85}{
\begin{tabular}{cc|cccccccc}
\toprule
 & $\ell$ & \multicolumn{8}{c}{$7$}\\
$\Gamma$ & \backslashbox{$I_z$}{$m$} & $-7$ & $-6$ & $-5$ & $-4$ & $-3$ & $-2$ & $-1$ & $0$\\
\midrule
$A_2$ & $2$ & & $\mathrm{i}\frac{1}{4}\sqrt{\frac{11}{3}}$ & & & & $\mathrm{i}\frac{1}{4}\sqrt{\frac{13}{3}}$ & \\
\midrule
\multirow{2}{0.25cm}{\centering{$E$}} & $0$ & & & & $\mathrm{i}\sqrt{\frac{1}{2}}$ & & & & \\
& $2$ & & $\mathrm{i}\frac{1}{4}\sqrt{\frac{13}{3}}$ & & & & $-\mathrm{i}\frac{1}{4}\sqrt{\frac{11}{3}}$ & &\\
\midrule
\multirow{6}{0.25cm}{\centering{$T_1$}} & $0$ & & & & $\frac{1}{4}\sqrt{\frac{7}{5}}$ & & & & $\frac{1}{2}\sqrt{\frac{33}{10}}$ \\
& $1$ &  $\frac{1}{8}\sqrt{\frac{65}{2}}$ & & & & $\frac{1}{8}\sqrt{\frac{77}{10}}$ & & \\
& $3$ & & & $\frac{1}{8}\sqrt{\frac{7}{10}}$ & & & & $\frac{1}{8}\sqrt{\frac{231}{10}}$ &  \\
& $0$ & & & & $\frac{1}{4}\sqrt{\frac{33}{5}}$ & & & & $-\frac{1}{2}\sqrt{\frac{7}{10}}$ \\
& $1$ & & & & & $\frac{1}{2}\sqrt{\frac{3}{10}}$ & & & \\
& $3$ & & & $\frac{1}{3}\sqrt{\frac{33}{10}}$ &  & & & -$\frac{1}{\sqrt{10}}$ &\\
\midrule
\multirow{6}{0.25cm}{\centering{$T_2$}} & $1$ & $\frac{3}{8}\sqrt{\frac{7}{2}}$ & & & & -$\frac{1}{24}\sqrt{\frac{143}{2}}$ & & &\\
& $2$ & & $\frac{\sqrt{2}}{24}$ & & & & $\frac{1}{24}\sqrt{\frac{143}{2}}$ & & \\
& $3$  & & & -$\frac{1}{24}\sqrt{\frac{13}{2}}$ & & & & -$\frac{1}{8}\sqrt{\frac{143}{6}}$ & \\
& $1$ & & & & & -$\frac{7}{12}\sqrt{2}$ & & & \\
& $2$ & & $\frac{1}{12}\sqrt{\frac{143}{2}}$ & & & & -$\frac{\sqrt{2}}{24}$ & & \\
& $3$  &  &  & $\frac{1}{6}\sqrt{\frac{11}{2}}$ & & & & $\frac{1}{\sqrt{6}}$ &\\
\bottomrule
\end{tabular}
}
\captionof{table}{Decomposition table for basis states of $\mathrm{SO}(3)$ with $\ell = 7$ and $m \leq 0$ into irreps of $\mathcal{O}$.}\label{T-A-09}
\end{center}

\begin{center}
\scalebox{0.80}{
\begin{tabular}{cc|ccccccc}
\toprule
 & $\ell$ & \multicolumn{7}{c}{$7$}\\
$\Gamma$ & \backslashbox{$I_z$}{$m$} & $1$ & $2$ & $3$ & $4$ & $5$ & $6$ & $7$\\
\midrule
$A_2$ & $2$ & & -$\mathrm{i}\frac{1}{4}\sqrt{\frac{13}{3}}$ & & & & -$\mathrm{i}\frac{1}{4}\sqrt{\frac{11}{3}}$\\
\midrule
\multirow{2}{0.25cm}{\centering{$E$}} & $0$ & & & & -$\mathrm{i}\sqrt{\frac{1}{2}}$ & & &\\
& $2$ & & $\mathrm{i}\frac{1}{4}\sqrt{\frac{11}{3}}$ & & & & -$\mathrm{i}\frac{1}{4}\sqrt{\frac{13}{3}}$ & \\
\midrule
\multirow{6}{0.25cm}{\centering{$T_1$}} & $0$ & & & & $\frac{1}{4}\sqrt{\frac{7}{5}}$ & & &\\
& $1$ & $\frac{1}{8}\sqrt{\frac{231}{10}}$ & & & & $\frac{1}{8}\sqrt{\frac{7}{10}}$ & &\\
& $3$ & & & $\frac{1}{8}\sqrt{\frac{77}{10}}$ & & & & $\frac{1}{8}\sqrt{\frac{65}{2}}$ \\
& $0$ & & & & $\frac{1}{4}\sqrt{\frac{33}{5}}$ & & & \\
& $1$ & -$\frac{1}{\sqrt{10}}$ & & & & $\frac{1}{2}\sqrt{\frac{33}{10}}$ & &\\
& $3$ & & & $\frac{1}{2}\sqrt{\frac{3}{10}}$ & & & & \\
\midrule
\multirow{6}{0.25cm}{\centering{$T_2$}} & $1$ & -$\frac{1}{8}\sqrt{\frac{143}{6}}$ & & & & -$\frac{1}{24}\sqrt{\frac{13}{2}}$ & &\\
& $2$  & & $\frac{1}{12}\sqrt{\frac{143}{2}}$ & & & & $\frac{\sqrt{2}}{24}$ & \\
& $3$  & & & -$\frac{1}{24}\sqrt{\frac{143}{2}}$ & & & & $\frac{3}{8}\sqrt{\frac{7}{2}}$\\
& $1$  & $\frac{1}{\sqrt{6}}$ & & & & $\frac{1}{6}\sqrt{\frac{11}{2}}$ & & \\
& $2$  & & -$\frac{\sqrt{2}}{24}$ & & & & $\frac{1}{12}\sqrt{\frac{143}{2}}$ & \\
& $3$  & & & -$\frac{7}{12}\sqrt{2}$ & & & &\\
\bottomrule
\end{tabular}
}
\captionof{table}{Decomposition table for basis states of $\mathrm{SO}(3)$ with $\ell = 7$ and $m >0$ into irreps of $\mathcal{O}$.}\label{T-A-10}
\end{center}
\smallskip
\begin{center}
\scalebox{0.75}{
\begin{tabular}{cc|ccccccccc}
\toprule
 & $\ell$ & \multicolumn{9}{c}{$8$}\\
$\Gamma$ & \backslashbox{$I_z$}{$m$} & $-8$ & $-7$ & $-6$ & $-5$ & $-4$ & $-3$ & $-2$ & $-1$ & $0$\\
\midrule
$A_1$ & $0$ & $\frac{1}{8}\sqrt{\frac{65}{6}}$ & & & & $\frac{1}{4}\sqrt{\frac{7}{6}}$ & & & & $\frac{\sqrt{33}}{8}$\\
\midrule
\multirow{4}{0.25cm}{\centering{$E$}} & $0$ & $\frac{1}{8}\sqrt{\frac{455}{246}}$ & & & & -$\frac{1}{4}\sqrt{\frac{41}{6}}$ & & & & $\frac{1}{8}\sqrt{\frac{231}{41}}$\\
& $2$ & & & $\frac{1}{4}\sqrt{\frac{273}{41}}$ & & & & -$\frac{1}{4}\sqrt{\frac{55}{41}}$ & & \\
& $0$ & $\frac{3}{2}\sqrt{\frac{11}{82}}$ & & & & & & & & -$\frac{1}{2}\sqrt{\frac{65}{41}}$\\
& $2$ & & & $\frac{1}{4}\sqrt{\frac{55}{41}}$ & & & & $\frac{1}{4}\sqrt{\frac{273}{41}}$ & & \\
\midrule
\multirow{6}{0.25cm}{\centering{$T_1$}} & $0$ & $\mathrm{i}\frac{1}{4\sqrt{57}}$ & & & & $\mathrm{i}\frac{1}{4}\sqrt{\frac{445}{57}}$\\
& $1$ &  & $\frac{1}{8}\sqrt{\frac{57}{2}}$ & & & & -$\frac{3}{8}\sqrt{\frac{91}{38}}$ & & & \\
& $3$ & & & & $\frac{5}{8}\sqrt{\frac{35}{114}}$ & & & & $\frac{1}{8}\sqrt{\frac{715}{114}}$ & \\
& $0$ & $\mathrm{i}\frac{1}{4}\sqrt{\frac{445}{57}}$ & & & & -$\mathrm{i}\frac{1}{4\sqrt{57}}$ & & & & \\
& $1$ & & &  & & & $\frac{3}{2}\sqrt{\frac{5}{38}}$ & & & \\
& $3$ & & & & $\frac{1}{2}\sqrt{\frac{13}{114}}$ &  & & & $\sqrt{\frac{77}{114}}$ \\
\midrule
\multirow{6}{0.25cm}{\centering{$T_2$}} & $1$ & & $\frac{1}{8}\sqrt{\frac{71}{2}}$ & & & & $\frac{3}{8}\sqrt{\frac{273}{142}}$ & & &\\
& $2$ & & & $\mathrm{i}\frac{3}{4}\sqrt{\frac{15}{142}}$ & & & & $\mathrm{i}\frac{1}{4}\sqrt{\frac{1001}{142}}$ & & \\
& $3$ & & & & -$\frac{5}{8}\sqrt{\frac{35}{142}}$ & & & & -$\frac{1}{8}\sqrt{\frac{715}{142}}$ & \\
& $1$ & & & & & & $\frac{1}{2}\sqrt{\frac{55}{142}}$ & & & \\
& $2$ & & & $\mathrm{i}\frac{1}{4}\sqrt{\frac{1001}{142}}$ & & & & -$\mathrm{i}\frac{3}{4}\sqrt{\frac{15}{142}}$ & & \\
& $3$  &  & &  & $\frac{1}{2}\sqrt{\frac{429}{142}}$ & & & & -$\sqrt{\frac{21}{142}}$ & \\
\bottomrule
\end{tabular}
}
\captionof{table}{Decomposition table for basis states of $\mathrm{SO}(3)$ with $\ell = 8$ and $m\leq 0$ into irreps of $\mathcal{O}$.}\label{T-A-11}
\end{center}

\begin{center}
\scalebox{0.71}{
\begin{tabular}{cc|cccccccc}
\toprule
 & $\ell$ & \multicolumn{8}{c}{$8$}\\
$\Gamma$ & \backslashbox{$I_z$}{$m$} & $1$ & $2$ & $3$ & $4$ & $5$ & $6$ & $7$ & $8$\\
\midrule
$A_1$ & $0$ & & & & $\frac{1}{4}\sqrt{\frac{7}{6}}$ & & & & $\frac{1}{8}\sqrt{\frac{65}{6}}$\\
\midrule
\multirow{4}{0.25cm}{\centering{$E$}} & $0$ & & & & -$\frac{1}{4}\sqrt{\frac{41}{6}}$ & & & & $\frac{1}{8}\sqrt{\frac{455}{246}}$\\
& $2$ & & -$\frac{1}{4}\sqrt{\frac{55}{41}}$ & & & & $\frac{1}{4}\sqrt{\frac{273}{41}}$ & & \\
& $0$ & & & & & & & & $\frac{3}{2}\sqrt{\frac{11}{82}}$\\
& $2$ & & $\frac{1}{4}\sqrt{\frac{273}{41}}$ & & & & $\frac{1}{4}\sqrt{\frac{55}{41}}$ & & \\
\midrule
\multirow{6}{0.25cm}{\centering{$T_1$}} & $0$ & & & & -$\mathrm{i}\frac{1}{4}\sqrt{\frac{445}{57}}$ & & & &-$\mathrm{i}\frac{1}{4\sqrt{57}}$ \\
& $1$ & $\frac{1}{8}\sqrt{\frac{715}{114}}$ & & & & $\frac{5}{8}\sqrt{\frac{35}{114}}$ & & &\\
& $3$ & & & -$\frac{3}{8}\sqrt{\frac{91}{38}}$ & & & & $\frac{1}{8}\sqrt{\frac{57}{2}}$ & \\
& $0$ & & & & $\mathrm{i}\frac{1}{4\sqrt{57}}$ & & & & -$\mathrm{i}\frac{1}{4}\sqrt{\frac{455}{57}}$\\
& $1$ & $\sqrt{\frac{77}{114}}$ & & & & $\frac{1}{2}\sqrt{\frac{13}{114}}$ & & &\\
& $3$ & & & $\frac{3}{2}\sqrt{\frac{5}{38}}$ & & & & \\
\midrule
\multirow{6}{0.25cm}{\centering{$T_2$}} & $1$ & -$\frac{1}{8}\sqrt{\frac{715}{142}}$ & & & & -$\frac{5}{8}\sqrt{\frac{35}{142}}$ & & &\\
& $2$ & & -$\mathrm{i}\frac{1}{4}\sqrt{\frac{1001}{142}}$ & & & & -$\mathrm{i}\frac{3}{4}\sqrt{\frac{15}{142}}$ & & \\
& $3$ & & & $\frac{3}{8}\sqrt{\frac{273}{142}}$ & & & & $\frac{1}{8}\sqrt{\frac{71}{2}}$ &\\
& $1$ & -$\sqrt{\frac{21}{142}}$ & & & & $\frac{1}{2}\sqrt{\frac{429}{142}}$ & & & \\
& $2$ & & $\mathrm{i}\frac{3}{4}\sqrt{\frac{15}{142}}$ & & & & -$\mathrm{i}\frac{1}{4}\sqrt{\frac{1001}{142}}$ & & \\
& $3$ & & & $\frac{1}{2}\sqrt{\frac{55}{142}}$ & & & & & \\
\bottomrule
\end{tabular}
}
\captionof{table}{Decomposition table for basis states of $\mathrm{SO}(3)$ with $\ell = 8$ and $m>0$  into irreps of $\mathcal{O}$.}\label{T-A-12}
\end{center}
\begin{center}
\scalebox{0.71}{
\begin{tabular}{cc|cccccccccc}
\toprule
 & $\ell$ & \multicolumn{10}{c}{$9$} \\
$\Gamma$ & \backslashbox{$I_z$}{$m$} & $-9$ & $-8$ & $-7$ & $-6$ & $-5$ & $-4$ & $-3$ & $-2$ & $-1$ & $0$\\
\midrule
$A_1$ & $0$ & & $\mathrm{i}\frac{1}{4}\sqrt{\frac{7}{3}}$ & & & & -$\mathrm{i}\frac{1}{4}\sqrt{\frac{17}{3}}$ & & & & \\
\midrule
$A_2$ & $2$ & & & & $\mathrm{i}\frac{1}{4}\sqrt{\frac{13}{2}}$ & & & & -$\mathrm{i}\frac{1}{4}\sqrt{\frac{3}{2}}$ & & \\
\midrule
\multirow{2}{0.25cm}{\centering{$E$}} & $0$ & $\mathrm{i}\frac{1}{4}\sqrt{\frac{17}{3}}$ & & & & $\mathrm{i}\frac{1}{4}\sqrt{\frac{7}{3}}$ & & & &\\
& $2$ & & & & $\mathrm{i}\frac{1}{4}\sqrt{\frac{3}{2}}$ & & & & $\mathrm{i}\frac{1}{4}\sqrt{\frac{13}{2}}$ & & \\
\midrule
\multirow{9}{0.25cm}{\centering{$T_1$}} & $0$ & & $\frac{7}{8}\sqrt{\frac{17}{690}}$ & & & & $\frac{1}{4}\sqrt{\frac{161}{30}}$ & & & & -$\frac{1}{8}\sqrt{\frac{429}{23}}$ \\
& $1$ & & & $\frac{1}{16}\sqrt{\frac{345}{2}}$ & & & & $\frac{3}{8}\sqrt{\frac{91}{230}}$ & & & \\
& $3$ & $\frac{1}{16}\sqrt{\frac{51}{230}}$ & & & & $\frac{41}{8\sqrt{138}}$ & & & & -$\frac{7}{16}\sqrt{\frac{143}{345}}$ & \\
& $0$ & & $\frac{29}{8}\sqrt{\frac{2}{53015}}$ & & & & $\frac{1}{2}\sqrt{\frac{2737}{4610}}$ & & & & $\frac{7}{4}\sqrt{\frac{2431}{10603}}$\\
& $1$ & & & & & & & $\frac{7}{8}\sqrt{\frac{9282}{53015}}$ & & & \\
& $3$ & $\frac{1}{2}\sqrt{\frac{461}{230}}$ & & & & $\frac{19}{4}\sqrt{\frac{17}{21206}}$ & & & & $\frac{11}{4}\sqrt{\frac{2431}{53015}}$ \\
& $0$ & & $\frac{1}{8}\sqrt{\frac{85085}{2766}}$ & & & & -$\frac{1}{4}\sqrt{\frac{715}{2766}}$ & & & & $\frac{3}{8}\sqrt{\frac{21}{461}}$ \\
& $1$ & & & & & & & -$3\sqrt{\frac{55}{922}}$ & & & \\
& $3$ & & & & & $\sqrt{\frac{1001}{2766}}$ & & & & $2\sqrt{\frac{35}{1383}}$ \\
\midrule
\multirow{6}{0.25cm}{\centering{$T_2$}} & $1$ & & & -$\frac{1}{16}\sqrt{\frac{3}{10}}$ & & & & -$\frac{1}{8}\sqrt{\frac{91}{10}}$ & & &\\
& $2$ & & & & $\frac{1}{4\sqrt{5}}$ & & & & $\frac{1}{4}\sqrt{\frac{39}{5}}$ & & \\
& $3$ & $\frac{1}{16}\sqrt{\frac{255}{2}}$ & & & & -$\frac{1}{8}\sqrt{\frac{3}{2}}$ & & & & -$\frac{1}{16}\sqrt{\frac{429}{5}}$ & \\
& $1$ & & & $\frac{1}{2}\sqrt{\frac{13}{10}}$ & & & & -$\frac{1}{4}\sqrt{\frac{21}{10}}$ & & & \\
& $2$ & & & & $\frac{1}{4}\sqrt{\frac{39}{5}}$ & & & & -$\frac{1}{4\sqrt{5}}$ & & \\
& $3$ & & & & & -$\frac{1}{4}\sqrt{\frac{13}{2}}$ & & & & $\frac{1}{4}\sqrt{\frac{11}{5}}$ & \\
\bottomrule
\end{tabular}
}
\captionof{table}{Decomposition table for basis states of $\mathrm{SO}(3)$ with $\ell = 9$ and $m\leq 0$ into irreps of $\mathcal{O}$.}\label{T-A-13}
\end{center}

\begin{center}
\scalebox{0.80}{
\begin{tabular}{cc|ccccccccc}
\toprule
 & $\ell$ & \multicolumn{9}{c}{$9$} \\
$\Gamma$ & \backslashbox{$I_z$}{$m$} & $1$ & $2$ & $3$ & $4$ & $5$ & $6$ & $7$ & $8$ & $9$\\
\midrule
$A_1$ & $0$ & & & & $\mathrm{i}\frac{1}{4}\sqrt{\frac{17}{3}}$ & & & & -$\mathrm{i}\frac{1}{4}\sqrt{\frac{7}{3}}$ & \\
\midrule
$A_2$ & $2$ & & & $\mathrm{i}\frac{1}{4}\sqrt{\frac{3}{2}}$ & & & & -$\mathrm{i}\frac{1}{4}\sqrt{\frac{13}{2}}$ & & \\
\midrule
\multirow{2}{0.25cm}{\centering{$E$}} & $0$ & & & & -$\mathrm{i}\frac{1}{4}\sqrt{\frac{7}{3}}$ & & & & -$\mathrm{i}\frac{1}{4}\sqrt{\frac{17}{3}}$ & \\
& $2$ & & -$\mathrm{i}\frac{1}{4}\sqrt{\frac{13}{2}}$ & & & & -$\mathrm{i}\frac{1}{4}\sqrt{\frac{3}{2}}$ & & & \\
\midrule
\multirow{9}{0.25cm}{\centering{$T_1$}} & $0$ & & & & $\frac{1}{4}\sqrt{\frac{161}{30}}$ & & & & $\frac{7}{8}\sqrt{\frac{17}{690}}$ & \\
& $1$ & -$\frac{7}{16}\sqrt{\frac{143}{345}}$ & & & & $\frac{41}{8\sqrt{138}}$ & & & & $\frac{1}{16}\sqrt{\frac{51}{230}}$ \\
& $3$ & & & $\frac{3}{8}\sqrt{\frac{91}{230}}$ & & & & $\frac{1}{16}\sqrt{\frac{345}{2}}$ & & \\
& $0$ & & & & $\frac{1}{2}\sqrt{\frac{2737}{4610}}$ & & & & $\frac{29}{8}\sqrt{\frac{2}{53015}}$ & \\
& $1$ & $\frac{11}{4}\sqrt{\frac{2431}{53015}}$ & & & & $\frac{19}{4}\sqrt{\frac{17}{21206}}$ & & & & $\frac{1}{2}\sqrt{\frac{461}{230}}$ \\
& $3$ & & & $\frac{7}{8}\sqrt{\frac{9282}{53015}}$ & & & & & & \\
& $0$ & & & & -$\frac{1}{4}\sqrt{\frac{715}{2766}}$ & & & & $\frac{1}{8}\sqrt{\frac{85085}{2766}}$ & \\
& $1$ & $2\sqrt{\frac{35}{1383}}$ & & & & $\sqrt{\frac{1001}{2766}}$ & & & & \\
& $3$ & & & -$3\sqrt{\frac{55}{922}}$ & & & & & & \\
\midrule
\multirow{6}{0.25cm}{\centering{$T_2$}} & $1$ & -$\frac{1}{16}\sqrt{\frac{429}{5}}$ & & & & -$\frac{1}{8}\sqrt{\frac{3}{2}}$ & & & & $\frac{1}{16}\sqrt{\frac{255}{2}}$ \\
& $2$ & & $\frac{1}{4}\sqrt{\frac{39}{5}}$ & & & & $\frac{1}{4\sqrt{5}}$ & & & \\
& $3$ & & & -$\frac{1}{8}\sqrt{\frac{91}{10}}$ & & & & -$\frac{1}{16}\sqrt{\frac{3}{10}}$ & & \\
& $1$ & $\frac{1}{4}\sqrt{\frac{11}{5}}$ & & & & -$\frac{1}{4}\sqrt{\frac{13}{2}}$ & & & & \\
& $2$ & & -$\frac{1}{4\sqrt{5}}$ & & & & $\frac{1}{4}\sqrt{\frac{39}{5}}$ & & & \\
& $3$ & & & -$\frac{1}{4}\sqrt{\frac{21}{10}}$ & & & & $\frac{1}{2}\sqrt{\frac{13}{10}}$ & & \\
\bottomrule
\end{tabular}
}
\captionof{table}{Decomposition table for basis states of $\mathrm{SO}(3)$ with $\ell = 9$ and $m>0$  into irreps of $\mathcal{O}$.}\label{T-A-14}
\end{center}

\end{appendices}




\renewcommand{\baselinestretch}{1.25} 



\end{document}